\documentclass[12pt,preprint]{aastex}

\newcommand{\HI}{\ion{H}{1}}

\begin{document}
\title{Tracing Cold \HI~Gas in Nearby, Low-Mass Galaxies}
\author{Steven R. Warren}
\affil{Minnesota Institute for Astrophysics, University of Minnesota, 116 Church St. SE, Minneapolis, MN 55455; 
warren@astro.umn.edu}

\author{Evan D. Skillman}
\affil{Minnesota Institute for Astrophysics, University of Minnesota, 116 Church St. SE, Minneapolis, MN 55455; 
skillman@astro.umn.edu}

\author{Adrienne M. Stilp}
\affil{Department of Astronomy, Box 351580, University of Washington, Seattle, WA 98195, USA; adrienne@astro.washington.edu}

\author{Julianne J. Dalcanton}
\affil{Department of Astronomy, Box 351580, University of Washington, Seattle, WA 98195, USA; jd@astro.washington.edu}

\author{J\"{u}rgen Ott}
\affil{National Radio Astronomy Observatory, 520 Edgemont Road, Charlotteville, VA 22903, USA; 
jott@nrao.edu }

\author{Fabian Walter}
\affil{Max Planck Institut f\"{u}r Astronomie, K\"{o}nigstuhl 17, D-69117 Heidelberg, Germany; walter@mpia.de}

\author{Eric A. Petersen}
\affil{Department of Physics, University of Illinois at Urbana-Champaign, 1110 West Green Street. Urbana, IL 61801;
eapeter2@illinois.edu}

\author{B\"{a}rbel Koribalski}
\affil{Australia Telescope National Facility, CSIRO Astronomy and Space Science,
PO Box 76, Epping NSW 1710, Australia; Baerbel.Koribalski@csiro.au}

\and

\author{Andrew A. West}
\affil{Department of Astronomy, Boston University, 725 Commonwealth Avenue,
Boston, MA 02215, USA; aawest@bu.edu}

\begin{abstract}

We analyze line-of-sight atomic hydrogen (\HI) line profiles of 31 nearby, low-mass galaxies selected from the 
Very Large Array - ACS Nearby Galaxy Survey Treasury (VLA-ANGST) and The \HI~Nearby Galaxy Survey (THINGS) to 
trace regions containing cold (T $\lesssim$ 1400 K) \HI~from observations with a uniform linear scale of 200 pc beam$^{-1}$.  
Our galaxy sample spans four orders of magnitude in total \HI~mass and nine magnitudes in M$_{B}$.  
We fit single and multiple component functions to each spectrum to isolate the cold, neutral medium given by 
a low dispersion ($<$6 km s$^{-1}$) component of the spectrum.  Most \HI~spectra are adequately 
fit by a single Gaussian with a dispersion of 8-12 km s$^{-1}$.  Cold \HI~is found in 23 of 27 ($\sim$85\%) galaxies after a
reduction of the sample size due to quality control cuts.  The cold \HI~contributes 
$\sim$20\% of the total line-of-sight flux when found with warm \HI.  Spectra best fit by a single Gaussian, but dominated by 
cold \HI~emission (i.e., have velocity dispersions $<$6 km s$^{-1}$) are found primarily beyond the optical radius of the 
host galaxy.  The 
cold \HI~is typically found in localized regions and is generally not coincident with the very highest surface density peaks 
of the global \HI~distribution (which are usually areas of recent star formation). 
We find a lower limit for the mass fraction of cold-to-total \HI~gas of only a few percent in each 
galaxy.

\end{abstract}  

\section{Introduction \label{intro}}

Dwarf irregular (dIrr) galaxies in the nearby universe are laboratories for studying the most 
fundamental properties of gas evolution and star formation.  Large, multi-wavelength studies of these systems
are only just beginning.  However, recent, high resolution surveys of large, spiral
galaxies in the ultraviolet 
(e.g., NGS - \citealt{gild07}), optical (e.g., NFGS - \citealt{jans00}; ANGST - \citealt{dalc09}), infrared 
(e.g., SINGS - \citealt{kenn03}), 
and radio (e.g., THINGS - \citealt{walt08}; HERACLES - \citealt{lero09}) have given us insight into where and when 
stars form in a galaxy.  They have also provided clues to the gas conditions surrounding sites of current 
star formation (\citealt{lero08}; \citealt{bigi08}).  The conclusion is that stars form in regions rich in cold, 
dense, molecular material \citep{kenn98, kenn12}.  These studies mainly focus on high metallicity systems, however.  How these results 
translate to low metallicity environments has 
not been fully explored, mainly due to the fact the most commonly observed tracer of molecular material, CO, is 
notoriously difficult to observe at low metallicity (e.g., \citealt{tayl98}; \citealt{baro00}; \citealt{lero05}; \citealt{schr12}).  

Nearby dIrr galaxies offer the opportunity to study the star formation process in low metallicity systems at high spatial
resolution.  Studies of the star formation rates in dIrr galaxies defined from UV, 24$\mu$m, and/or H$\alpha$ emission (e.g., 
\citealt{lero08}; \citealt{bigi08}; \citealt{royc09}) have found a non-linear correlation with the local atomic hydrogen (\HI) 
distributions.  However, the dependence
of star formation on molecular gas content is apparent in the large spiral galaxies observed in \citet{lero08} and 
\citet{bigi08}.  The star formation rate surface density is shown to correlate linearly with the molecular gas surface
density in these galaxies.  These results can be understood if one assumes that efficient star formation requires molecular 
hydrogen (e.g., \citealt{krum09} and references therein).  

Based on theoretical studies, the very character of the interstellar medium (ISM) is likely to change at low 
metallicity (e.g., \citealt{spaa97}).  \citet{glov11} modeled the so called X-factor which measures the relationship between
amount of detected CO emission and the abundance of molecular hydrogen.  They found that the amount of CO drops substantially at
low metallicities, consistent with prior predictions (e.g., \citealt{malo88}).  This deficit of CO at low metallicities is supported by 
observations showing relatively low or no detections of 
CO emission in nearby dIrr galaxies (e.g., \citealt{tayl98, baro00, lero05, schr12}).  Consequently, this lack of metals also results in lower
amounts of polycyclic aromatic hydrocarbon (PAH) emission (e.g., \citealt{engl05, jack06}). The PAHs are important to the formation of 
the cold ISM since they are critical in the shielding of UV and soft X-ray photons.    

The long standing problem of studying the molecular component in the low metallicity dwarfs continues.  
We expect dIrr galaxies to contain molecular hydrogen since they {\it are} forming stars.  However, since direct observations of 
the molecular gas responsible for forming stars in large samples of dIrr 
galaxies via CO are currently not feasible, we must find a different tracer of star-forming gas. 

One intriguing idea is to find the immediate precursors of the molecular gas.  The assembly of star forming molecular clouds is 
generally believed to require cold \HI~(e.g., \citealt{wolf03}; \citealt{krum09}).  In our galaxy, cold \HI~clouds have been 
observed to surround and even intermix with molecular clouds (e.g., \citealt{krco10}) through studies of \HI~narrow 
self absorption (HINSA; \citealt{li03}).  One promising technique to find cold \HI~gas was pioneered by 
\citet{youn96, youn97} in a sample of nearby, star forming dIrr galaxies.  
These authors used high angular and spectral resolution \HI~data of two nearby dIrr galaxies to 
decompose line-of-sight spectra into narrow and broad Gaussian components.  The narrow-lined gas ($\sigma$ $\sim$ 4.5 
km s$^{-1}$) was found only in specific regions within the \HI~disk while the broad-lined gas ($\sigma$ $\sim$ 10 km 
s$^{-1}$) was found along every line-of-sight observed.  The narrow Gaussian component was 
attributed to cold (T$\lesssim$1000 K) \HI~gas while the broad lined gas was assumed to be warm, neutral hydrogen 
(T$\gtrsim$5000 K).  \citet{youn03} studied three more dIrr galaxies, finding that they, too, contained evidence for cold
\HI~gas.  Other authors have used this technique to discover cold \HI~in a small number of other galaxies.  
\citet{debl06} investigated the \HI~distribution in NGC 6822 and also found regions rich with cold 
\HI~gas.  Recent work by \citet{begu06} found evidence of cold \HI~gas in six dIrr galaxies from 
the Faint Irregular Galaxies GMRT Survey (FIGGS; \citealt{begu08}).  To date, cold \HI~in emission has been discovered in 
10 nearby dIrr galaxies.  

Within the limited data available, different measurements of the cold \HI~do appear to give comparable results.  Two galaxies, 
DDO 210 and GR8, from the sample of \citet{youn03} overlapped with that of \citet{begu06}.  
\citet{youn03} used the Very Large Array (VLA) at a linear scale of $\sim$200 pc beam$^{-1}$ with a velocity resolution of 1.3 km s$^{-1}$ 
while \citet{begu06} used the Giant Metrewave Radio Telescope (GMRT) at a linear resolution of 
$\sim$300 pc beam$^{-1}$ with a velocity resolution of 1.65 km s$^{-1}$.  Despite the use of different facilities and different spatial/spectral
resolutions, both studies find similar results for both the spatial distributions and velocity dispersions of the narrow component.  This agreement 
between the two studies suggests that the measurement of cold \HI~is generally robust, and not extremely 
sensitive to the exact observing parameters.

The technique of identifying a cold neutral medium has also been used in nearby, high metallicity spiral galaxies.  \citet{brau97} 
isolated cold \HI~gas using relatively low spectral resolution (6 km s$^{-1}$) imaging from the VLA for 11 of the closest spiral galaxies.  
He combined spectra from discrete radial bins and found that these combined spectra all had a narrow 
Gaussian core (FWHM $\lesssim$ 6 km s$^{-1}$) superposed onto broad (FWHM $\sim$ 30 km s$^{-1}$) Lorentzian wings.  He also 
found that the cold \HI~gas is filamentary and is found in clumps, preferentially in the spiral arms.  Since the majority of 
star formation in spiral galaxies occurs in molecular clouds within the spiral arms (e.g., \citealt{gord04}), it is not surprising 
to find the bulk of the cold \HI~associated with these features. 

The purpose of our study is to build upon the previous results of \citet{youn96,youn97}, \citet{youn03}, \citet{begu06},
and \citet{debl06} in order to provide limits to the locations and amount of cold \HI~gas in a large sample of 31 nearby,
low-mass galaxies using a common spatial resolution.  We describe our galaxy sample in \S\ref{obs} and our methods of signal 
extraction in \S\ref{idents}.  Our results are described in \S\ref{results} and we end with a summary of our conclusions in \S\ref{conclusions}.

\section{Galaxy Sample, Observations, and Data Reduction \label{obs}}

Galaxies for this work were taken from two major \HI~surveys of nearby galaxies: The \HI~Nearby Galaxy
Survey (THINGS; AW0605; \citealt{walt08}) and the Very Large Array - Advanced Camera for Surveys Nearby Galaxy Survey Treasury 
(VLA-ANGST; AO0215; Ott et al. {\it in preparation}).  Our sample consists of 24 galaxies from VLA-ANGST and 7 dwarf galaxies in 
the M81 Group from THINGS.  The VLA-ANGST observations all have a high spectral
resolution of 0.65 - 2.6 km s$^{-1}$.  The 7 galaxies from the THINGS sample have velocity resolutions of 1.3 - 2.6 km s$^{-1}$.  
The high velocity resolution is crucial for detecting narrow velocity
components.  Other galaxies from the THINGS sample were not used because they had velocity
resolutions of 5 km s$^{-1}$, too coarse for this type of analysis. Our final sample contains 31 galaxies with
distances ranging from 1.3 - 5.3 Mpc (average 2.9 Mpc) and \HI~masses between
4$\times$10$^{5}$ - 1$\times$10$^{9}$ $M_{\odot}$.

All observations were made using the Very Large Array\footnotemark\footnotetext{The VLA telescope of the National 
Radio Astronomy Observatory is operated by Associated Universities, Inc. under a cooperative
agreement with the National Science Foundation.} (VLA) observatory.  We briefly outline the basic 
reduction procedure here, but note that both surveys followed similar recipes, described fully in \citet{walt08} 
and Ott et al. 
({\it in preparation}).  Standard {\sc AIPS}\footnotemark\footnotetext{The Astronomical Image Processing System
(AIPS) has been developed by the NRAO.} processing of spectral line data was followed.  Phase, bandpass, and
flux corrections were made using typical VLA calibrators.  Two sets of data cubes were produced: a natural weighted
cube with a typical beam size of $\sim$10$\arcsec$ and a robust weighted cube with a beam size of $\sim$7$\arcsec$. 
Both sets of data have rms noise values of $\sim$1 mJy beam$^{-1}$ in a single line-free channel.  We use the robust 
weighted cubes in this work, which allows us to use the smallest possible uniform linear scale.  Further, robust weighting produces 
minimal side-lobes allowing for a clean-beam which better approximates the dirty beam structure than that of the more complex structure seen
in the dirty-beam of natural weighting \citep{brig95}.  The better dirty-beam approximation of the robust weighting is ideal for the 
extended emission observed in \HI~data.

The main difference between the data for the two surveys is due to the inclusion of updated 
Expanded VLA (EVLA) antennas into the array for the VLA-ANGST survey (the EVLA has since been renamed the ``Karl G.
Janksy Very Large Array").  Unfortunately, the conversion of the digital EVLA signal to
an analog signal (to be compatible with the VLA signal) aliased power into the first 0.5 MHz of the baseband.
This only affected the EVLA-EVLA baselines and they were subsequently removed from the data (see Ott et al. {\it in preparation} for full
details). 
Aliasing did not correlate into the mixed VLA-EVLA baselines and the VLA-VLA baselines were unaffected.  To compensate for the loss
of the EVLA-EVLA baselines, additional observation time was spent on each source.  Table \ref{galprop} gives an 
overview of the major observational properties of our galaxy sample.  The table columns are: 1) galaxy name, 2,3) RA and
DEC, 4) distance in Mpc (from TRGB measurements in \citealt{dalc09} unless otherwise noted), 5) total \HI~mass, 6) absolute 
$B$-band magnitude, 7) the major-axis length in arcminutes at the $\sim$25 mag arcsec$^{-2}$ level (except for KKH 98 and
KK 230, which use the Holmberg system of $\sim$26.5 mag arcsec$^{-2}$; \citealt{kara04}), 8) inclination \citep{kara04}, 9)
the 200 pc beam size (see \S\ref{methods}), 10) the spectral resolution, and 11) the rms of the noise
in the line-free channels of the data cube.  While some metallicity estimates exist for the most massive galaxies in our sample, most do not have 
measured abundances.  Using the relation in \citet{lee06} which relates M$_{B}$ to the oxygen
abundance (see also \citealt{berg12}), we expect a range of 7.0$\lesssim$12+log(O/H)$\lesssim$8.2 with a median value of 7.7 
for the low mass galaxies in our sample.  

\section{Identifying Narrow Spectral Components \label{idents}}
\subsection{Methodology \label{methods}}

We analyze line-of-sight 
spectra following methods similar to those outlined in \citet{youn96,youn97} and \citet{youn03}.  First, we smoothed
each image cube with a Gaussian kernel to a common linear scale of 200 pc beam$^{-1}$.  To do this we computed the angular size equivalent
of 200 pc for each galaxy using the distances given in Table \ref{galprop}.  We then used the CONVL task in AIPS which incorporates the initial
beam size, beam position angle, and requested output beam size to compute the required convolving kernel.  We note that our circular
beam sizes can sample slightly larger linear scales depending on galaxy inclination.
Spectra were then extracted through every 1$\farcs$5 pixel (1$\arcsec$ pixels for NGC~247 and NGC~3109).  Extracting spectra from every pixel 
over-samples the beam, but for our purposes, does not influence the results.  Also, the beam
sizes reported in Table \ref{galprop} and the typical galaxy rotation speeds of 10-20 km s$^{-1}$ are small enough that kinematic
broadening is minimal, regardless of galaxy inclination.  Note that spectral broadening would work to hide a narrow 
component, not create one. 

Each extracted spectrum was successively fitted using our own routines with two different functional forms: a single Gaussian and a double 
Gaussian.  We also attempted to fit the spectra with a fourth order Gauss-Hermite polynomial of the form
\begin{equation} 
\phi(x) = ae^{-y^{2}/2}\left[1+\frac{h_{3}}{\sqrt{6}}(2\sqrt{2}y^{3}-3\sqrt{2}y)
+\frac{h_{4}}{\sqrt{24}}(4y^{4}-12y^{2}+3)\right]
\end{equation}  
where $y\equiv(x-b)/\sigma_{herm}$ \citep{vand93}, $h_{3}$
and $h_{4}$ measure the amplitudes of an asymmetric and a symmetric deviation from an underlying Gaussian profile with amplitude $a$,
central velocity $b$, and standard deviation $\sigma_{herm}$.  Gauss-Hermite polynomials offer useful characterizations of
non-Gaussian line profiles, but as \citet{youn03} point out, the direct 
relationship between the Gauss-Hermite variables and the physical conditions of the gas is not obvious.  In general, the results of the
double Gaussian and Gauss-Hermite polynomial fits were in excellent agreement, although the detection efficiency of the
Gauss-Hermite polynomials is much lower.  However, we expect the observed spectra to have a Gaussian 
profile if the gas has an isothermal density distribution \citep{merr93}. 
Since the relationship between the Gauss-Hermite polynomial parameters and the
physical nature of the gas is not obvious, we only report the results of the single and double Gaussian fits.

A spectrum was identified as containing narrow-line \HI~only if multiple criteria were fulfilled.  First, we only 
fit extracted spectra with a S/N greater than 10 (see \S\ref{sims}).  We define S/N in the same manner as \citet{youn96}, that is, the 
peak in the spectrum divided by the rms noise in the line free channels of the data cube.  We also required each individual component of the double 
Gaussian fit to have a minimum S/N of 3.1.  This minimum S/N requirement was also used by \citet{youn96}, \citet{youn97}, and 
\citet{youn03} and is motivated by our desire to obtain significant detections.  Also, each spectrum (single or multiple component) was 
required to have a
velocity dispersion greater than the velocity resolution of the data (0.65, 1.3, or 2.6 km s$^{-1}$) beyond the 1$\sigma$ 
errors in the fits.  Requiring the velocity dispersion plus error to be greater than the velocity resolution ensures we are not
fitting noise spikes in the data.  Fortunately, none of the fitted spectra were affected by this criterion.  Typical errors in the velocity 
dispersion of each single and double Gaussian component are roughly 0.7 km s$^{-1}$.  We also do not enforce any restrictions on the central 
velocity offsets between individual components, leaving this as a free parameter.

To quantify which function best fitted each spectrum, we first computed the variance of the residuals for each fit.  
The ratio of the variances between the single and double Gaussian fits were then compared statistically via a single-tailed
F-test.  A double Gaussian fit was determined to be statistically significant if the probability of improvement 
over a single Gaussian fit was 95\% or greater.  We were as conservative with our approach as possible in order to 
provide secure lower limits to the total amount of narrow-line \HI~in each galaxy.  \citet{youn96,youn97}, \citet{youn03}, and
\citet{begu06} were less conservative in their approach, requiring only a 90\% significance of improvement over a single 
Gaussian profile.  Because of our relatively high S/N cut-off, the results of our fitting are relatively independent 
of the F-test criterion that we use, as long as it is above roughly 70\%.  Since the number of detections of double Gaussian profiles increases 
very slowly as the cut-off is reduced from 95\% down to 70\%, we chose 95\% confidence to establish a very firm lower limit on the total cold 
\HI~content.  Establishing a realistic upper limit is more problematic and is highly dependent upon assumptions.  

We further adopt the criteria of \citet{debl06} that cold \HI~has a velocity dispersion of 
less than 6 km s$^{-1}$.  A velocity dispersion of 6 km s$^{-1}$ results in an upper limit to the gas temperature of $\sim$1400 K
(assuming (3/2)kT = (1/2)$m_{p}$$\sigma$$^{2}$).  Thus, we did not ascribe a second 
component as cold \HI~when the fits detected multiple components with similar velocity dispersions between 6-10 km
s$^{-1}$ or one component with a velocity dispersion of 6-10 km s$^{-1}$ and another component with a much higher dispersion.  The vast majority of 
spectra that were
best fit by double Gaussian profiles with both velocity dispersions greater than 6 km s$^{-1}$ were double peaked (in both low and high inclination
galaxies) which is normally associated with expanding structures (e.g., \citealt{brin86}).  Furthermore, these higher velocity dispersions 
are not typical for the cold \HI~described in the
literature with velocity dispersions of 2-6 km s$^{-1}$ (\citealt{youn96, brau97, youn97, youn03, debl06, begu06}); we therefore rejected these
locations as having narrow line emission.  \citet{debl06} also noted that the
bulk of the spectra within this category for NGC~6822 were double-peaked, leading them to make a cut at 6 km s$^{-1}$.  Thus, having a cut 
at 6 km s$^{-1}$ ensures that we are both reporting cold gas and eliminating a large fraction of the double peaked spectra.  The 
double-peaked spectra eliminated by this criteria represent less than 1\% of the total spectra and are not included in any of the following 
analyses.  We also identify
spectra that were sufficiently fit by a single Gaussian profile with a velocity dispersion of less than 6 km s$^{-1}$ as cold \HI.
Limiting our cold \HI~identifications to 
gas with velocity dispersions less than 6 km s$^{-1}$ ensures we are not misidentifying warm \HI~as cold \HI.

One possible scenario is that the broad component is the combination of multiple cold gas (narrow Gaussian) clouds.  Given that the typical
line-of-sight spectrum is best fit by a single Gaussian profile (see \S\ref{results}), the multiple cold \HI~clouds would have to conspire in such a 
way to maintain this Gaussianity.  While this scenario is physically possible, it requires a large number of
    cold HI clouds with distributions in velocities and amplitudes that would 
    systematically combine to maintain a total Gaussian velocity dispersion 
    equal to the observed velocity dispersion for the warm HI phase.  High
    resolution absorption line studies would be required to distinguish
    whether the broad component is truly broad or a composite of many narrow
    components.  However, if we assume that the 
galaxies can be approximated by a thin disk we would
expect that the line centroid of the cold \HI~to be at or very near the line centroid of the warm \HI.  Warm \HI~gas scale heights in 
dwarf galaxies tend to be a few hundred pc (e.g., \citealt{warr11}) compared to their few kpc diameters, so this is not a bad assumption.  Offset velocity gas 
due to inclination effects would then contribute to the broadening of the warm (broad Gaussian) component and less so to the cold (narrow) component 
gas.  Not surprisingly, the average central velocity offset between the components of the best fit double Gaussian profiles in our sample is 
0.35 km s$^{-1}$, which is less than the velocity resolution of our data.  Thus, the observations are most simply interpreted as a combination of cold 
gas near the central velocity (when observed) and nearly ubiquitous warm gas.  We find it highly unlikely that the broad component is a combination of 
multiple narrow components.  This leads us to the conclusion that the \HI~is dominated by the warm phase and that the cold phase represents a much
smaller fraction of the ISM in our sample of dwarf galaxies.

\begin{figure}
\includegraphics[width=175mm]{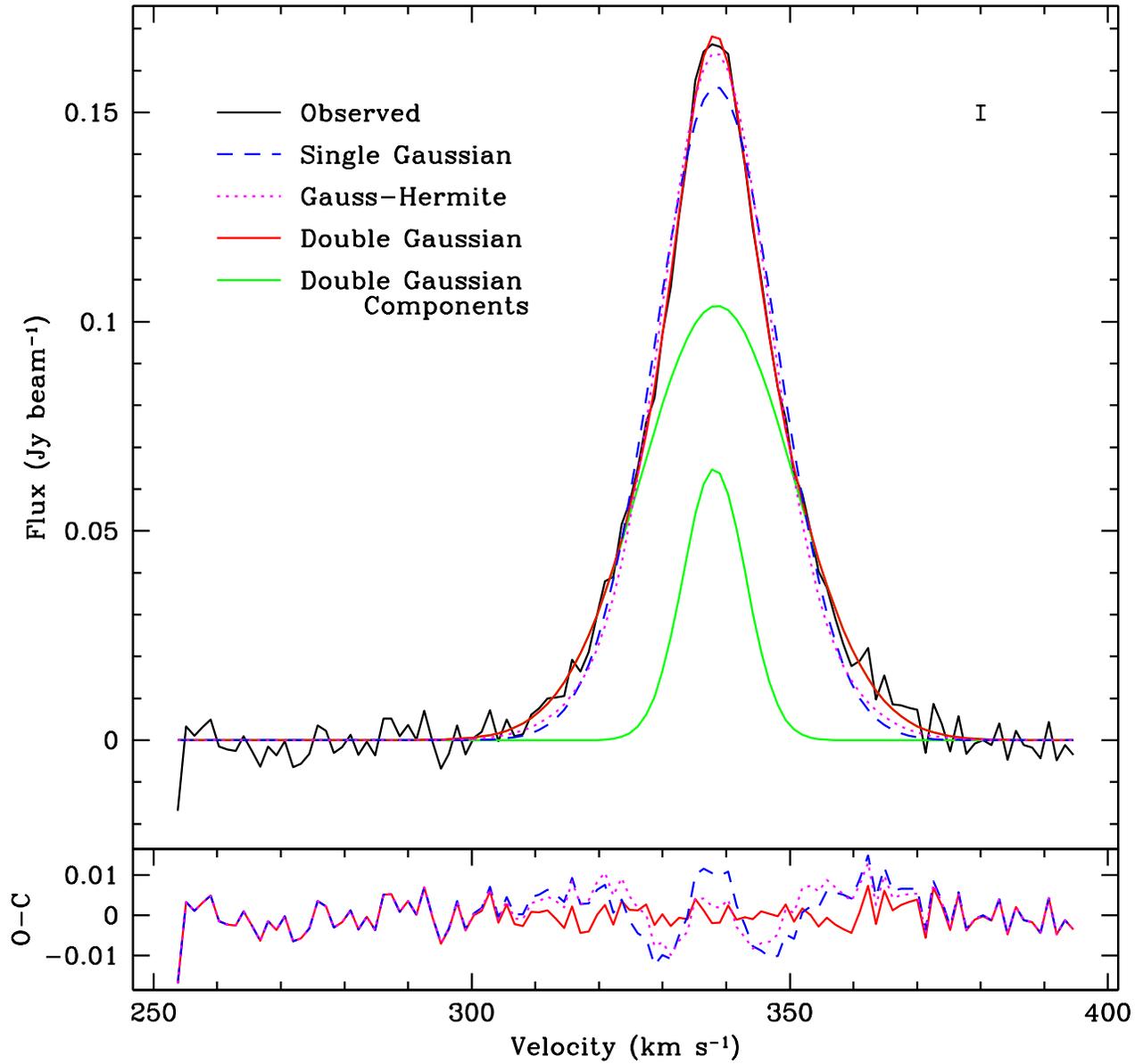}
\caption{An example spectrum from Sextans A.  The error on the points is shown in the upper right corner.  This spectrum
has a S/N of 37 and is clearly not fit well by a single Gaussian profile.  The residuals to the fits are shown in the bottom
panel.  Both the double Gaussian profile and Gauss-Hermite polynomial
statistically fit the observed spectrum better than the single Gaussian profile at the 99.9\% confidence level in a
single-tailed F-test.
\label{spectrum}}
\end{figure}

The top panel of Figure \ref{spectrum} shows an extracted spectrum 
(black) from the VLA-ANGST observations of Sextans A with single Gaussian profile (blue dashed), double Gaussian profile (solid red), 
and Gauss-Hermite polynomial fits (dotted magenta).  The
bottom panel shows the residual to the fit.  This spectrum has a clear narrow peak that is not fit well by the single
Gaussian profile.  Both the double Gaussian profile and the Gauss-Hermite polynomial fit the spectrum statistically better than the single
Gaussian profile at the 99.9\% confidence level.  Although the Gauss-Hermite polynomial is a better fit than the single Gaussian
profile, the residuals clearly show that the double Gaussian profile is even better.

In summary, we identify cold \HI~by the following process: 1) only spectra with a S/N greater than 10 
were fit, 2) each spectrum was successively fitted with a single and double Gaussian function, 3) each individual component of the double 
Gaussian fit was required to have a minimum S/N of 3.1, 4) the velocity dispersion plus error had to be greater than the velocity 
resolution of the data, 5) double Gaussian fits were accepted if they improved the fit over a single Gaussian profile at the 95\% or 
greater confidence level in a single tailed F-test, and 6) cold \HI~can be described by single or double Gaussian profiles 
with velocity dispersions less than 6 km s$^{-1}$.

\subsection{Simulations of Synthetic Spectra \label{sims}}

There are two factors which could lead to non-detections of existing cold \HI.  Cold 
\HI~could be missed simply because the column density is too low to reach our S/N requirement, or it could be missed because a higher S/N is 
required to deblend the cold \HI~from the warm \HI.  As a test of our ability to extract multiple components from a given spectrum, we produced 
three sets of 50,000 uniformly distributed synthetic spectra that spanned the S/N of our sample (10 $\leq$ S/N $\leq$ 60; one set each at 0.65, 
1.3, and 2.6 km s$^{-1}$ velocity resolution).  Each spectrum contained 
two Gaussian profiles that are representative of the warm and cold \HI~gas;
a ``broad" Gaussian with $\sigma_{broad}$ between 8 and 12 km s$^{-1}$ and a ``narrow" Gaussian with $\sigma_{narrow}$ between
3 and 6 km s$^{-1}$.  We allowed for an offset of $\pm$2.5 km s$^{-1}$ between the line centroids.  Velocity offsets larger than 2.5 km s$^{-1}$
allow for much higher detection fractions since the spectra become more asymmetric.  The observed cold \HI~line centroids are also very similar to
the warm \HI~line centroids (as discussed in \S\ref{methods}).  For each S/N, we randomized the 
amplitude of the cold \HI~Gaussian.  We then chose the warm \HI~Gaussian such that the S/N was conserved. As for the analysis with our
observations, each individual
component was required to have a S/N $>$ 3.1. A new set of Gaussians was produced if each component did not have a S/N $>$
3.1.  We next added random noise to each spectrum and then passed each randomly generated spectrum through our fitting routine to 
gauge our recovery rate.

\begin{figure}
\includegraphics[width=175mm]{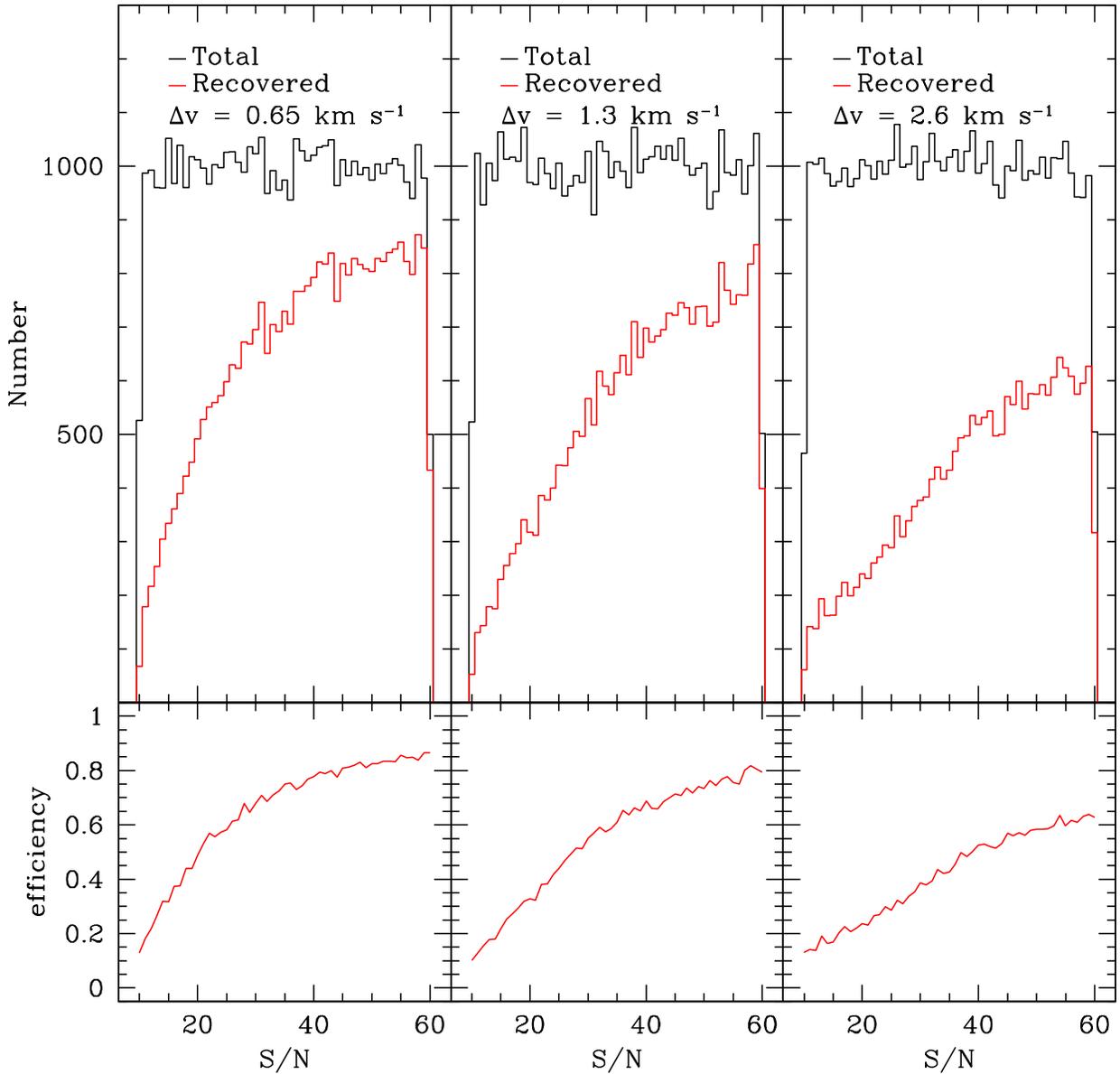}
\caption{Results from the synthetic spectra containing two Gaussian components randomly generated and processed with our 
fitting routine.  {\it Top Row:} The total sample (black) along with the recovered profiles (red; includes best-fit single 
Gaussian profiles with a velocity dispersion less than 6 km s$^{-1}$ for the 
three different velocity resolutions (0.65, 1.3, and 2.6 km s$^{-1}$).  
{\it Bottom Row:}  Our recovery efficiency versus S/N for the double Gaussian profiles.
\label{mcarlo}}
\end{figure}

The top row of Figure \ref{mcarlo} shows histograms of the total number of synthetic spectra (solid black line) and the total 
number identified as requiring multiple components by a 
double Gaussian fit (solid red line) as a function of S/N.  The red histograms also include the locations best fit by a single Gaussian with a velocity
dispersion of less than 6 km s$^{-1}$.  The bottom row shows the detection efficiency as a function of S/N for the 
double Gaussian fits (including the best fit single Gaussians with a velocity
dispersion of less than 6 km s$^{-1}$).  Figure \ref{mcarlo} shows that we never identify all double Gaussian profiles 
within the range of S/N values scrutinized and our recovery efficiency worsens as the velocity resolution decreases.  The differences in recovery
efficiency between the 0.65 and 1.3 km s$^{-1}$ simulations are less pronounced than from 1.3 to 2.6 km s$^{-1}$.

Table \ref{simtab} shows the results of the simulations for each velocity resolution for the double Gaussian profiles that were identified.  Column (1) 
is the velocity resolution, column (2) is the ratio between the input and extracted broad Gaussian amplitude (A$_{b,sim}$/A$_{b,extr}$), columns (3) and (4) 
are the average differences between the simulated and extracted broad Gaussian central velocities ($\Delta$v$_{broad}$) and velocity dispersions 
($\Delta$$\sigma$$_{broad}$), column is the ratio between the input and extracted narrow Gaussian amplitude (A$_{n,sim}$/A$_{n,extr}$), and columns
(6) and (7) are the average differences between the simulated and extracted narrow Gaussian central velocities ($\Delta$v$_{nar}$) and velocity dispersions 
($\Delta$$\sigma$$_{nar}$), respectively.  These results show that
when we do recover two Gaussian profiles, our routines accurately reproduce the input Gaussian parameters.

\begin{figure}
\includegraphics[width=175mm]{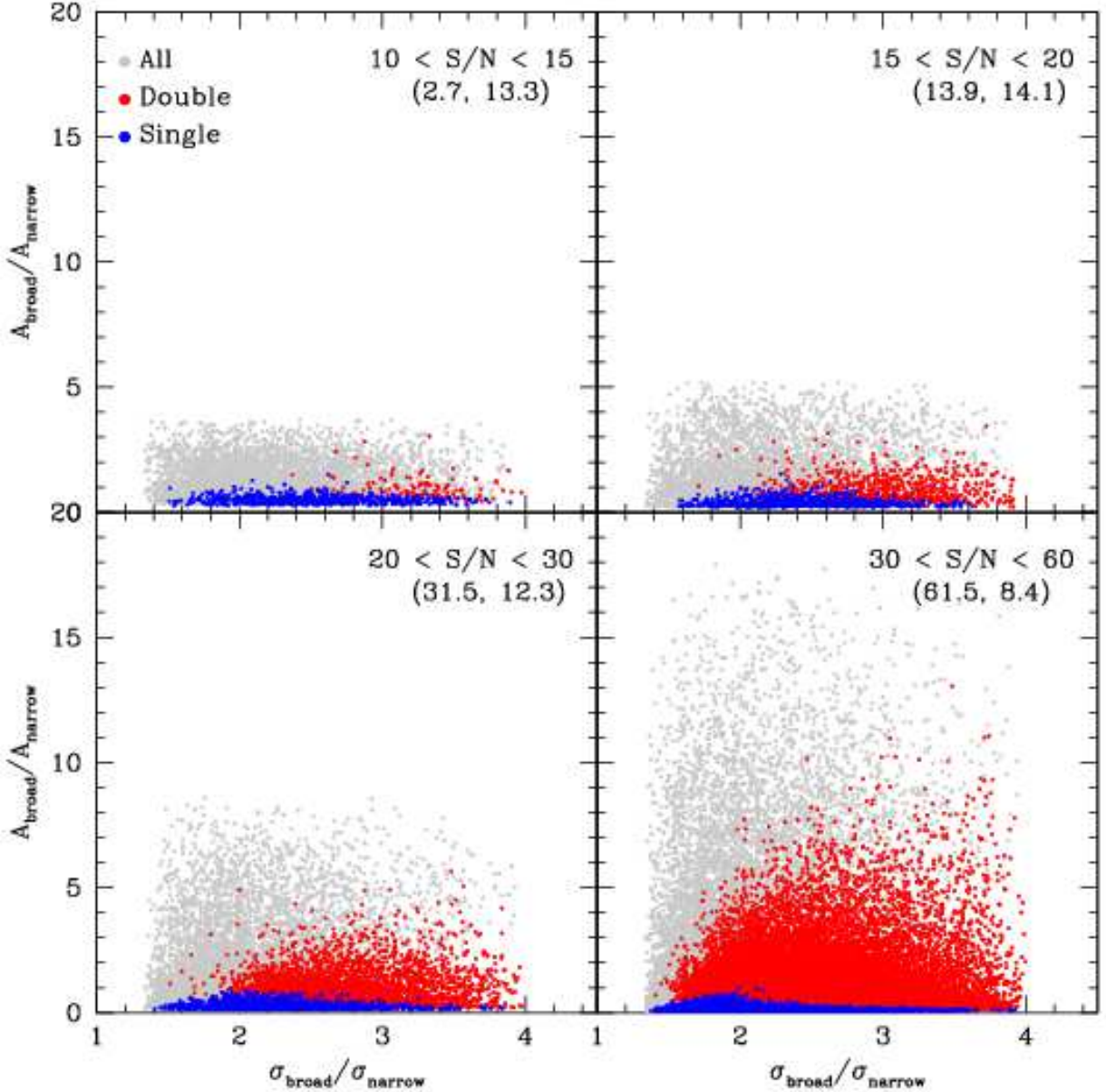}
\caption{The ratio of the broad and narrow Gaussian amplitudes (A$_{\mathrm{broad}}$/A$_{\mathrm{narrow}}$) as a function of 
the ratio of the broad and narrow velocity dispersions ($\sigma_{\mathrm{broad}}$/$\sigma_{\mathrm{narrow}}$) for the 1.3 km 
s$^{-1}$ velocity resolution simulation in four different S/N bins.  The grey
dots are all of the simulated spectra.  The red dots 
are those spectra identified as containing two components by our fitting routines, and the blue dots are the simulated spectra 
that were identified only containing a single Gaussian component with a velocity dispersion of less than 6 km
s$^{-1}$.  The values in the parentheses are the recovery percentages for the best fit double (left) and single
(right) Gaussians.  We clearly 
identify more of the spectra at higher S/N values.  Our routines have a harder time recovering the two Gaussian components when 
the velocity dispersions are similar.
\label{mcarlo2}}
\end{figure}

In Figure \ref{mcarlo2} we plot the amplitude ratio of the broad and narrow Gaussian components (A$_{\mathrm{broad}}$/A$_{\mathrm{narrow}}$) versus the 
ratio of the
velocity dispersions ($\sigma_{\mathrm{broad}}$/$\sigma_{\mathrm{narrow}}$) for the 1.3 km s$^{-1}$ velocity resolution simulation.  We show these ratios
in four different S/N bins to understand where our fitting routines have trouble identifying the two Gaussian components.  The grey points represent all of the
simulated spectra, the red points are those spectra identified as containing two Gaussian components, and the blue points are the simulated spectra best
fit by a single Gaussian with a velocity dispersion of less than 6 km s$^{-1}$ (even though the input spectra contains both a broad and narrow Gaussian
as described above).  As is shown in Figure \ref{mcarlo}, the recovery efficiency is dependent upon the S/N. The best fit single Gaussians are
predominately in a regime where the narrow component dominates the spectra (amplitude ratios less than 1 and velocity dispersion ratios less than 2.5). 
We also have difficulty identifying the narrow component when the broad component dominates the spectra in all S/N bins.  Lastly, our routines do not
pick up multiple components when the velocity dispersions are similar.  These plots show some of the complexities in simulating and recovering 
multiple Gaussian components.  When we do identify the spectra as containing
multiple components, however, our routines accurately reproduce the input parameters.  The simulations indicate that 
we can be confident of our detections in the observed data.  However, the simulations also show that we can only reliably compute 
lower limits to the cold \HI~content.  A complete census of the cold \HI~would require both higher S/N spectra and a detailed knowledge of 
the intrinsic distributions in A$_{\mathrm{broad}}$/A$_{\mathrm{narrow}}$ and $\sigma_{\mathrm{broad}}$/$\sigma_{\mathrm{narrow}}$ (in order to make 
incompleteness corrections).

We also produced 50,000 synthetic spectra of single Gaussian profiles to quantify our false positives.  These
spectra were generated using a velocity dispersion between 8 and 12 km s$^{-1}$ with S/N values between 10 and 60.  We used a velocity resolution of 2.6
km s$^{-1}$ since this gives the worst recovery results.  With our
acceptance criteria, we failed to find a single false identification of multiple components in our synthetic spectra.  
We do not start to see the potential for false detections until the confidence level is lowered to $\sim$70\%.  
These simulations give us confidence that our fitting results are robust to false detections.   

\section{Results \label{results}}

\subsection{Cold \HI~Detections \label{colddets}}

Each galaxy in our sample varies in terms of the total number of spectra and observed minimum, maximum, and average S/N 
(and column density).  In total, we have analyzed roughly 4,100 independent lines-of-sight.  
Table \ref{fitres} summarizes the observed spectral properties of each galaxy.  The columns indicate 1)
galaxy name, 2) total number of independent lines-of-sight scrutinized (N$_{t}$), 3) average S/N of the extracted spectra 
($<$S/N$>_{t}$), 4) the approximate column density at a S/N of 10 (N$_{\mathrm{HI},min}$), 5) the peak column density 
(N$_{\mathrm{HI},peak}$), and 6) the average column density ($<$N$_{\mathrm{HI}}$$>$).  All of the \HI~spectra in DDO 82, 
KDG 73, KK 230, and KKH 98 have S/N values below our threshold and as a result had no spectra analyzed, reducing our final 
sample to 27 galaxies.

Table \ref{indres} summarizes the results of our fitting
routine.  The columns are described as follows: 1) galaxy name, 2) the average S/N of the spectra where cold \HI~is found 
($<$S/N$>_c$), 3) the areal filling factor of the cold \HI~defined as the ratio
of the number of cold \HI~detections and the total number of scrutinized spectra ($\mathcal{F}_{fill}$), 4) the 
cold-to-total \HI~mass fraction defined as the ratio of the total summed column densities of the cold \HI~Gaussian profiles 
and the total summed column densities in the areas scrutinized ($\mathcal{F}_{mass}^{low}$), 5) an estimate of the upper limit to the 
cold-to-total \HI~mass fraction assuming each line-of-sight contains cold \HI~($\mathcal{F}_{mass}^{up}$; see \S\ref{amfrac}), 
6 \& 7) the average velocity dispersion 
of the narrow and broad Gaussian components ($<$$\sigma_{n}$$>$ and $<$$\sigma_{b}$$>$), 
and 7) the average velocity dispersion of the locations where a single Gaussian profile was sufficient to describe the 
spectrum ($<$$\sigma_{s}$$>$).  Column 5 also includes all of the best fit single Gaussian profiles with a
velocity dispersion of less than 6 km s$^{-1}$, while column 7 excludes them.  

We detected cold \HI~in 23 out of the 27 galaxies in our final sample.  Figure \ref{2plots} shows the 
spatial distribution of the cold \HI~for each galaxy.  The left panels show the total
integrated \HI~intensity maps with contours of 10$^{20}$ and 10$^{21}$ cm$^{-2}$ overlaid.  The right panels have the \HI~surface
density contours overlaid on the locations of the cold \HI~emission.  The majority of the cold \HI~is clumped in localized regions.  
Typically, the cold \HI~is not spatially coincident with the very highest peaks in the total \HI~distribution in a given galaxy, although it is mainly 
concentrated in locations where the total \HI~column density exceeds the canonical threshold for star formation of 
$\sim$10$^{21}$ cm$^{-2}$ \citep{skil87}.  Cold \HI~described in each of the previous studies also shows a preference for being located
in regions of high column density, but not necessarily associated with the highest \HI~columns.  The concentration of the cold \HI~in 
these regions is not unexpected given these location have the highest S/N values.  The four galaxies in
which we do not detect cold 
\HI~(DDO~99, MCG09-20-131, NGC~3741, and UGCA~292) show no distinguishing characteristics to give us insight as to why they are non-detections.
Each of their total \HI~masses, M$_{B}$ values, distances, stellar disk sizes, and star formation rates (see \S\ref{h2sfr}) are similar to 
other galaxies in our sample that do show cold \HI~detections.

\subsection{Comparison with Previous Work}

Fortunately, our sample of 27 galaxies had some overlap with previous studies using similar methods to detect cold \HI. 
\citet{youn03} detected cold \HI~in UGCA 292 and GR8 using the VLA while \citet{begu06} detected it in DDO~53, MCG09-20-131, GR8 and 
M81~Dwarf~A using the GMRT.  These five galaxies are amongst the faintest and smallest galaxies in our sample and as a result, we 
detected cold \HI~in only 3 of these 5 galaxies: DDO~53, GR8, and M81~Dwarf~A.  Our selection criteria 
failed to produce a cold \HI~signature in UGCA~292 or MCG09-20-131.

The three galaxies in which we did detect cold \HI~are included in Figure \ref{2plots}.  For GR8, \citet{youn03} and 
\citet{begu06} each found cold \HI~in both the northern and south-western portions of the galaxy while \citet{youn03} also 
found some evidence in the eastern portion of the galaxy.  In contrast, we only found cold \HI~in
the northern region of GR8.  For M81~Dwarf~A we detect cold \HI~only in the eastern side of the galaxy while \citet{begu06} find a
hint of cold \HI~in the western side of the galaxy as well.  For DDO~53 we found good overall agreement with the previous study, 
although our total detection area is smaller.  

The data quality between each survey is similar, yet they provide minor differences in their results.  Our simulations suggest velocity resolution 
differences may play a role in the detection differences between the studies.  The velocity resolutions in our sample for these galaxies 
are 0.65, 1.3, and 2.6 km 
s$^{-1}$ for GR8, M81~Dwarf~A, and DDO~53, respectively.  The literature velocity resolutions are 1.3 
and 1.65 km s$^{-1}$ for GR8, 1.65 km s$^{-1}$ for M81~Dwarf~A, and 1.65 km s$^{-1}$ for DDO~53.  The galaxy with the most similar cold 
\HI~distribution, DDO~53, has the 
worst velocity resolution in our sample, 0.85 km s$^{-1}$ worse than the literature value.  GR8 has been analyzed with three different
velocity resolutions and each study finds slightly different results.  We reduced our velocity resolution to 1.3 km s$^{-1}$ for GR8 to
compare to the other studies and obtained no difference in our results.  M81~Dwarf~A was analyzed with similar velocity resolutions, yet
there still exists slight differences.  Despite analyzing data from different observatories with different velocity resolutions, the general results are all
similar.  It seems likely that the differences in each galaxy arise from 
small differences in the selection criteria and beam shape.  For example, both \citet{youn03} and \citet{begu06} used naturally weighted data cubes which, 
on average, produce higher S/N spectra.  However, as noted in
\S\ref{obs}, the robust weighting for combined array data is ideal for studying extended emission due to the better clean beam approximation of the
dirty beam in the CLEAN algorithm.  The previous studies also used a cutoff 
of 90\% in the F-test statistic and also allow the narrow Gaussian component to be larger than 6 km s$^{-1}$, finding values up to $\sim$8 km
s$^{-1}$.   

\subsection{Are the Cold \HI~Non-Detections Significant? \label{nodets}}

\begin{figure}
\includegraphics[width=175mm]{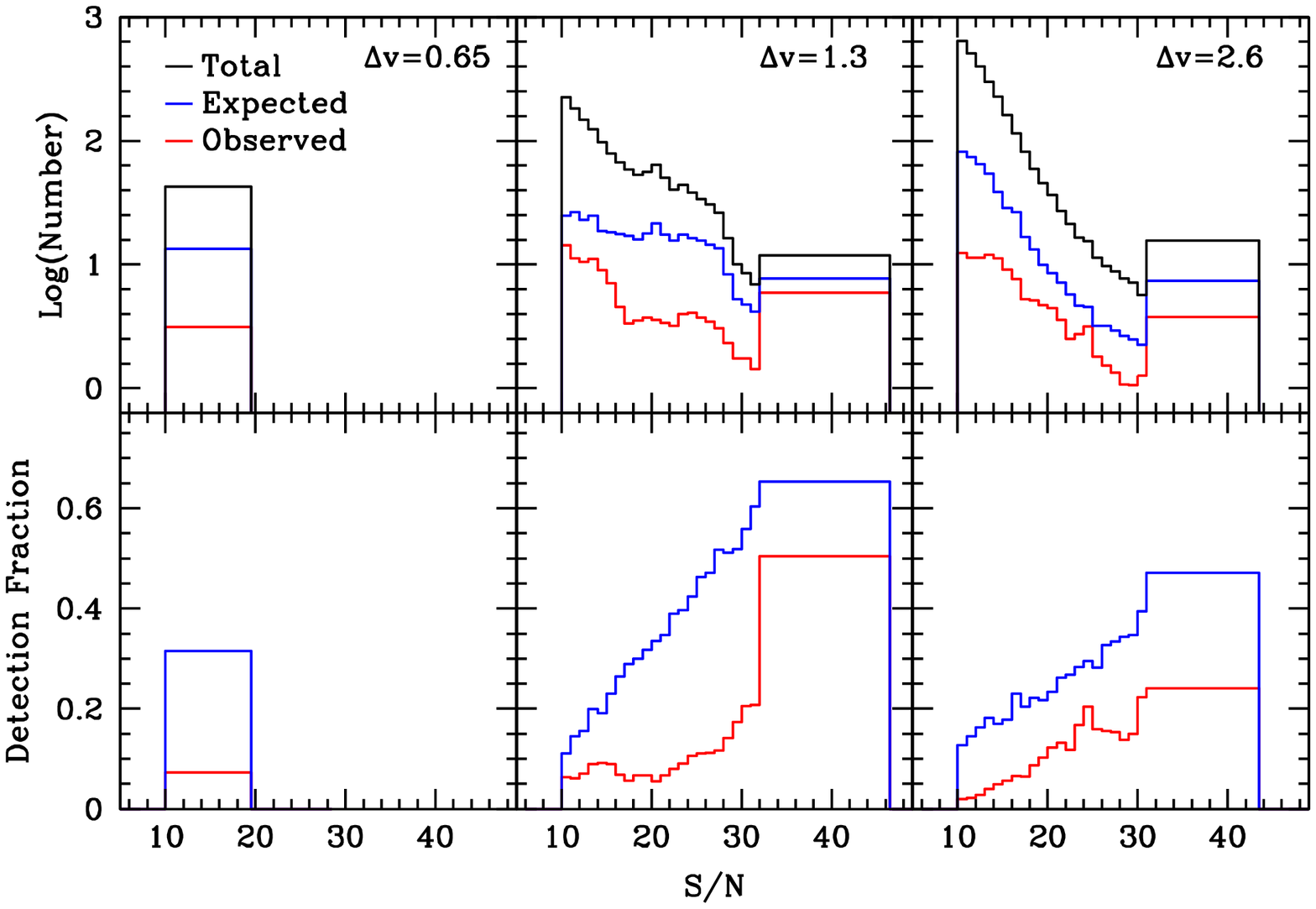}
\caption{{\it Top:} Histograms of the total observed sample for all galaxies (black), the expected number of cold \HI~detections given our
detection efficiency and assuming every line-of-sight contained multiple components (blue), and the actual number of cold \HI~detections
(red) as a function of S/N for each velocity resolution.  The histogram bins were chosen such that each 
contains at least one independent line-of-sight.  The large gap between the blue and red histograms demonstrates the cold \HI~identified with this technique
is not ubiquitous. {\it Bottom:} The detection fraction as a function of S/N. 
\label{expected}}
\end{figure}

If each spectrum consisted of both a broad and narrow component and were not inherently a single Gaussian profile, we would 
expect to find far more locations with a narrow line signature than we actually do.  The top panel of Figure \ref{expected} shows 
a histogram for every independent line-of-sight (black line), the expected number of
identified narrow line detections (blue) (based upon our detection efficiency as calculated in \S\ref{sims} and the assumption 
that a narrow component exists at 
each location), and the actual number of narrow line detections (red) as a function of S/N.  The bottom panel shows the detection 
fraction as a function of S/N.  Figure \ref{expected} demonstrates the large gap between the 
number of spectra for which we are sensitive to the presence of narrow \HI~and the number of spectra for which there are 
detections.  Our expected recovery fraction is insensitive to small changes in the reasonable ranges used for the parameters. 

The lower S/N spectra have fewer fractional detections.  As discussed in 
\S\ref{sims}, the lower fractional detections at lower S/N values are due, in part, to our recovery efficiency but may also be due 
to a minimum total \HI~column density required for the appearance of cold \HI.  \citet{kane11} recently observed \HI~in the Milky 
Way and determined a minimum column density threshold of 2$\times$10$^{20}$ cm$^{-2}$ for the formation of a cold phase of \HI, 
which is just at the column density limit of where our sample begins.  The top row of Figure \ref{cdplot} shows histograms of all 
of the observed column densities in our galaxy sample (black), and the column
densities where we detect cold \HI~(red).  The bottom row shows our detection fraction as a function of column density.

The small fraction of narrow line detections at high S/N compared to what would be expected clearly shows that the cold \HI~that can be 
identified with this technique is inconsistent with a ubiquitous distribution.  It should also be noted that there 
could exist cold \HI~at every line-of-sight below a S/N of 3.1 that we would be 
insensitive to.  However, it would be unphysical for the presence of gas to be related to our S/N criterion and not to the 
\HI~column density.  Furthermore, absorption line studies in the Milky way indicate that cold \HI~gas is not ubiquitous
(e.g., \citealt{begu10}; \citealt{kane11}).

\subsection{The Velocity Dispersions of Cold and Warm \HI}

\begin{figure}
\includegraphics[width=175mm]{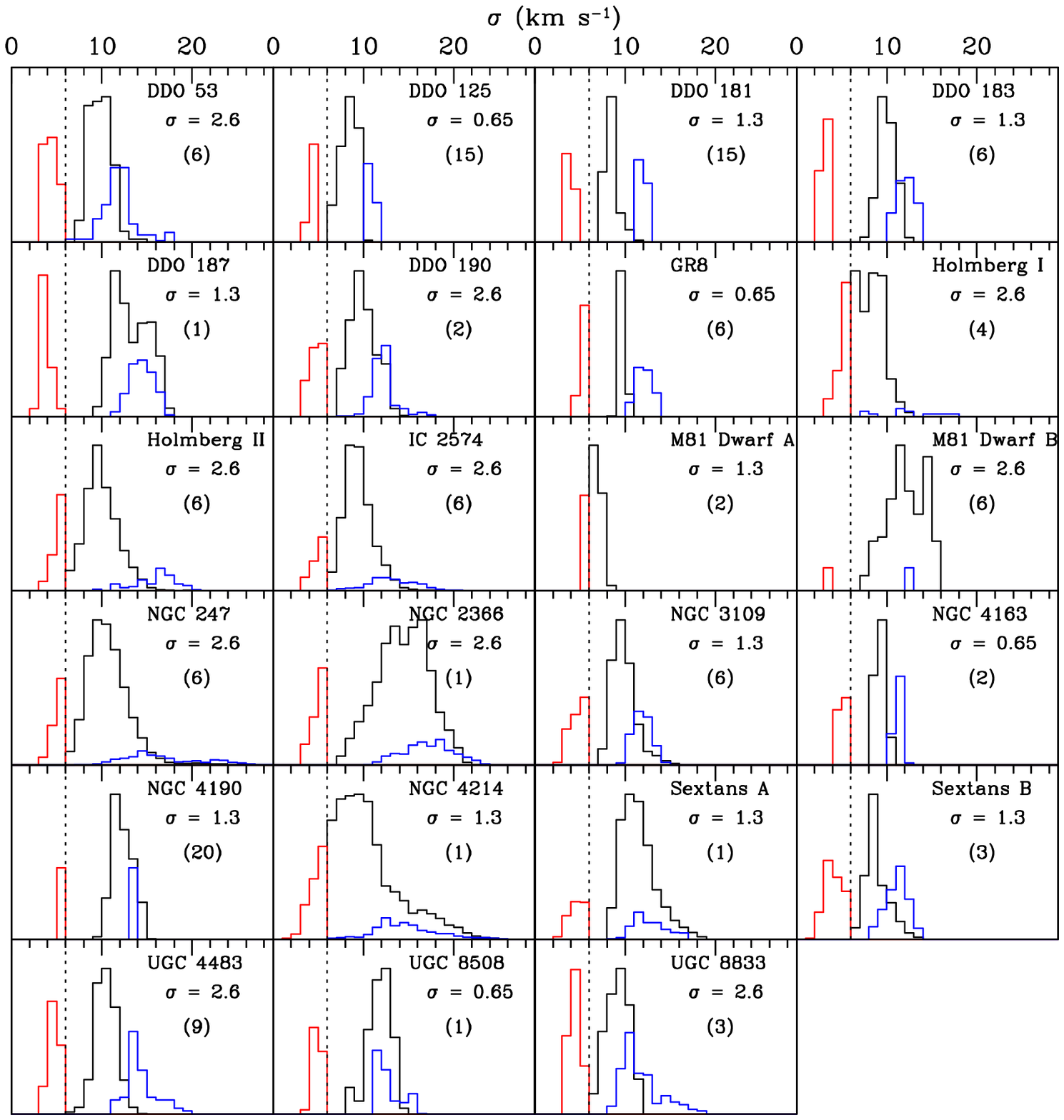}
\caption{Histograms of the velocity dispersions of the best fit single Gaussian profiles (black), narrow component (red), and 
broad component (blue).  The single Gaussian profile distributions have been scaled down by the number in
the parentheses in order to discern the other two
components.  The dashed vertical line denotes our narrow-line cutoff of 6 km s$^{-1}$.  The velocity resolution ($\sigma$) is listed
below the galaxy names.  Typically, the broad component 
velocity dispersions overlap with the single Gaussian profile velocity dispersions indicating they arise from the same gas phase.
\label{veldisp}}
\end{figure}

The typical velocity dispersion for the cold \HI~is $\sim$4.5 km s$^{-1}$.  The broad components and best-fit
single Gaussian profiles vary by galaxy, but they generally have similar values.  We expect the broad and best fit single
components to be similar if they are both tracing the same warm \HI.  Figure \ref{veldisp} shows histograms of the narrow
(red), broad (blue), and single (black) velocity dispersions for each galaxy.  The dotted vertical line denotes
our cold \HI~cutoff limit of 6 km s$^{-1}$.  The single component histograms have been scaled in order to
show the narrow and broad component histograms more clearly.  Generally, the velocity dispersions of the broad components 
overlap with those of the single components.  However, the peak in the broad component is offset from the peak in the 
single component towards higher values.  This behavior is also seen for the majority of the cases in the literature, and is
probably due to the fact that it is easier to identify a narrow component when the broad component has a much larger 
value.

\begin{figure}
\includegraphics[width=175mm]{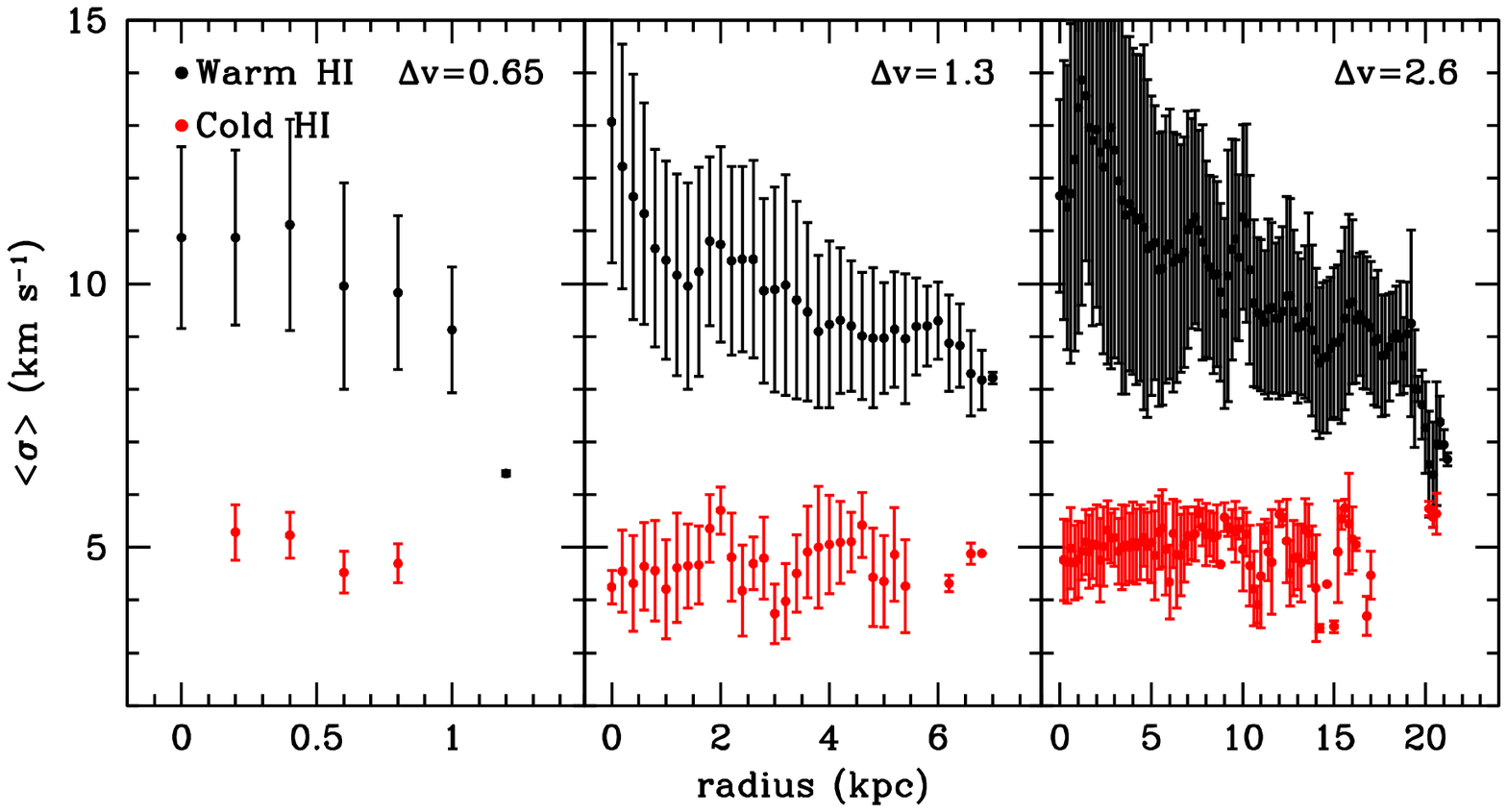}
\caption{The average velocity dispersion, $<$$\sigma$$>$, %
as a function of radius for the warm (black) and cold (red) \HI~gas for
each of our velocity resolutions.  We
have omitted the locations best fit by a single Gaussian with a velocity dispersion of less than 6 km s$^{-1}$.  The averages 
are taken over 200 pc bins (the beam size). The error bars are the dispersions of the values in the bins.  The velocity of the warm 
component decreases with radius indicating a possible decrease in turbulence as the radius increases from the main stellar body.
Some of the errorbars at large radii for the 2.6 km s$^{-1}$ velocity resolution overlap with values
below our 6 km s$^{-1}$ cutoff.  This is due to the dispersions being affected by a few larger values
in these bins.
\label{disprad}}
\end{figure}

In Figure \ref{disprad}, we plot the average velocity dispersion, $<$$\sigma$$>$, of the warm (black) and cold (red) \HI~gas as a
function of radius for each of our different velocity resolutions.  We have omitted the locations best fit by a single Gaussian
with a velocity dispersion of less than 6 km s$^{-1}$.  The average warm \HI~velocity dispersion declines with radius.  This
trend is similar to the total \HI~velocity dispersion versus radius seen by \citet{tamb09}.  The decrease in the velocity
dispersion with increasing radius may indicate a decrease in turbulent energy supplied by the underlying
stellar population as the radius increases away from the center of the stellar disk.

\subsection{The Areal and Mass Fractions of Cold \HI \label{amfrac}}

Recent work on galaxy simulations of spiral disks have put limits on the predicted volume filling factor of the different gas phases 
in the ISM (e.g., \citealt{deav04}).  The cold (T $<$ 1000 K) ISM occupies $\lesssim$20\% of a
galaxy's volume, depending upon the star formation (supernova) rate.  The cold neutral medium, however, may
occupy only a few percent of the volume \citep{mcke77}.  The actual value will vary by galaxy and must also be sensitive to the available
ISM coolants (e.g., C$+$, CO, dust, etc.) and the global gravitational potential.

Without knowing the exact 3-dimensional structure of
the galaxies in our sample, we cannot calculate the volume filling fractions.  Instead, we calculate the areal filling
fractions ($\mathcal{F}_{fill}$), i.e., the ratio of the number of lines-of-sight with detected cold \HI~and the total number of
lines-of-sight with S/N $>$ 10 (see Table \ref{indres}, column 3).  We derive an average $\mathcal{F}_{fill}$ value of 9\%.  The true volume 
filling fractions based upon our detections are most likely even lower than these values since the scale height of the warm \HI~will be larger 
than that of the cold \HI.  The low filling fractions we compute are consistent with the ISM model of \citet{mcke77}.

\begin{figure}
\includegraphics[width=175mm]{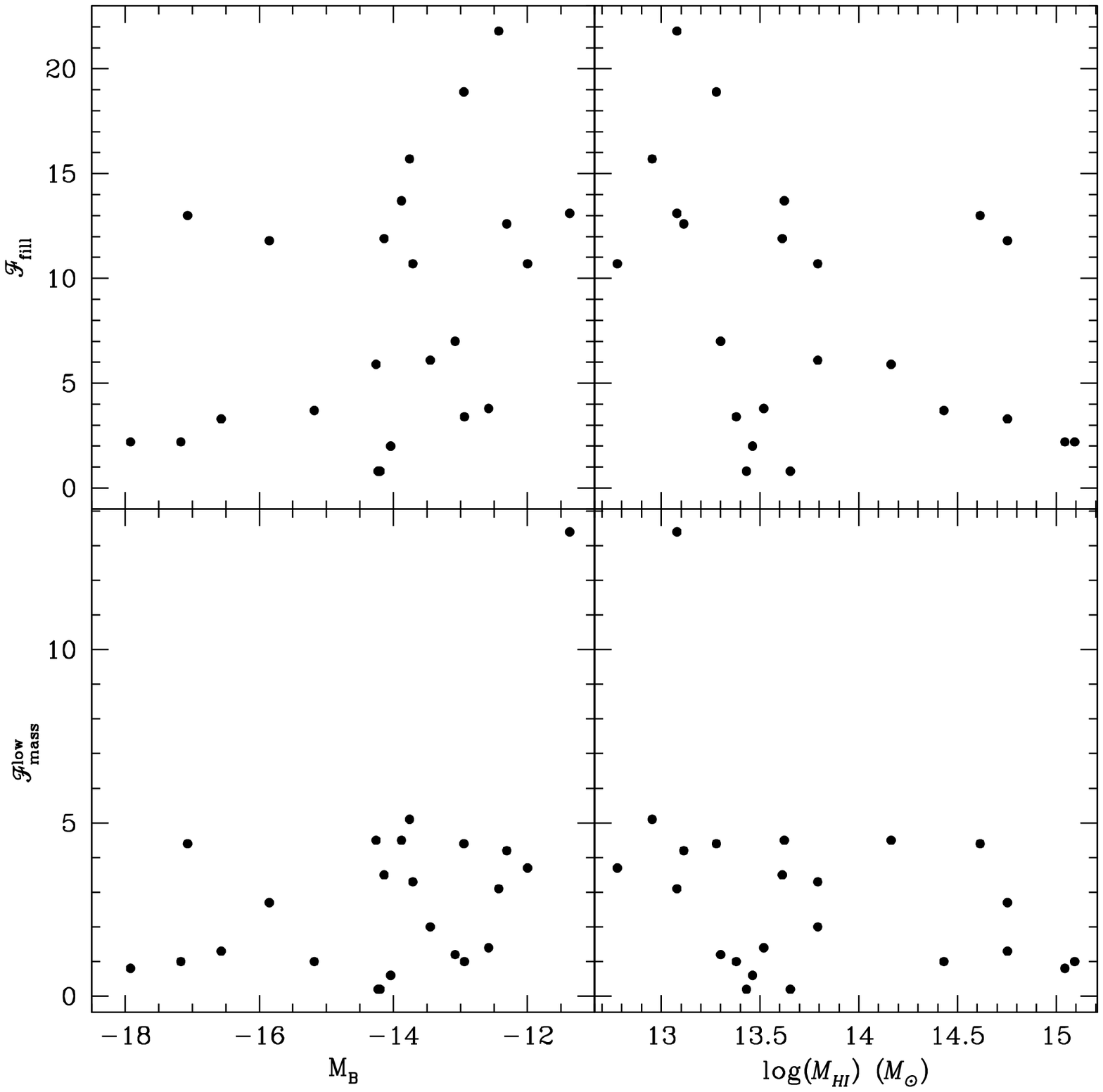}
\caption{The areal fraction for the cold gas ($\mathcal{F}_{fill}$) and the lower limit to the fraction
of \HI~gas in the cold phase ($\mathcal{F}_{mass}^{low}$) as a function M$_{B}$ (left) and $M_{HI}$ (right).  The areal filling 
fraction and mass fractions do not correlate with M$_{B}$ or $M_{HI}$.
\label{fracs}}
\end{figure}

We also computed a lower limit to the total mass (flux) contribution of the cold \HI~for each galaxy using the narrow component of the Gaussian
fits ($\mathcal{F}_{mass}^{low}$; see Table \ref{indres}, column 4).  \citet{youn96,youn97} and \citet{debl06} found that roughly 20\% 
of the total \HI~mass 
is in the cold \HI, which is significantly higher than the $\sim$3\% typical
for our values of $\mathcal{F}^{low}_{mass}$.  It is unclear as to the reason for this large discrepancy since the data are of
    similar quality.  The biggest differences may lie in the higher S/N resulting from the natural weighting
    these other studies use as compared to the robust weighting we employ. 
    Also, \citeauthor{debl06} use a different statistical test in defining the
    acceptance of cold \HI~detections.  These authors accept a double Gaussian
    fit as statistically better than a single Gaussian fit if the $\chi$$^{2}$ value is
    improved by 10\%.  Furthermore, our values are lower limits given our selection criteria and sensitivities.  
Figure \ref{fracs} plots $\mathcal{F}_{fill}$ and $\mathcal{F}_{mass}^{low}$ as a function of M$_{B}$ (left) and $M_{HI}$ (right).  No
obvious trends arise between the areal filling fraction and mass fraction with absolute $B$-band magnitude or total \HI~mass.

If we assume (as noted in \S\ref{nodets}) that each line of sight has cold \HI~just below our detection limits, we can 
estimate an upper limit to the amount of cold \HI~in each galaxy.  To do this we have used a
representative cold \HI~Gaussian of amplitude 3.1$\sigma$ and velocity dispersion of 4.5 km s$^{-1}$ at each location where we do not 
detect cold \HI.  We then add this ``extra" mass to our detected cold \HI~mass to produce an upper limit to the cold \HI~mass
fraction ($\mathcal{F}_{mass}^{up}$; see column 5 in 
Table \ref{indres}).  These values of $\mathcal{F}^{up}_{mass}$ can range up to 50\%.  However, since 
the cold \HI~is most frequently associated with higher values of N$_{\mathrm{HI}}$ (see 
\S\ref{colddets}), it is unlikely that the true values of cold \HI~mass get as high as
these upper limits.

\begin{figure}
\includegraphics[width=175mm]{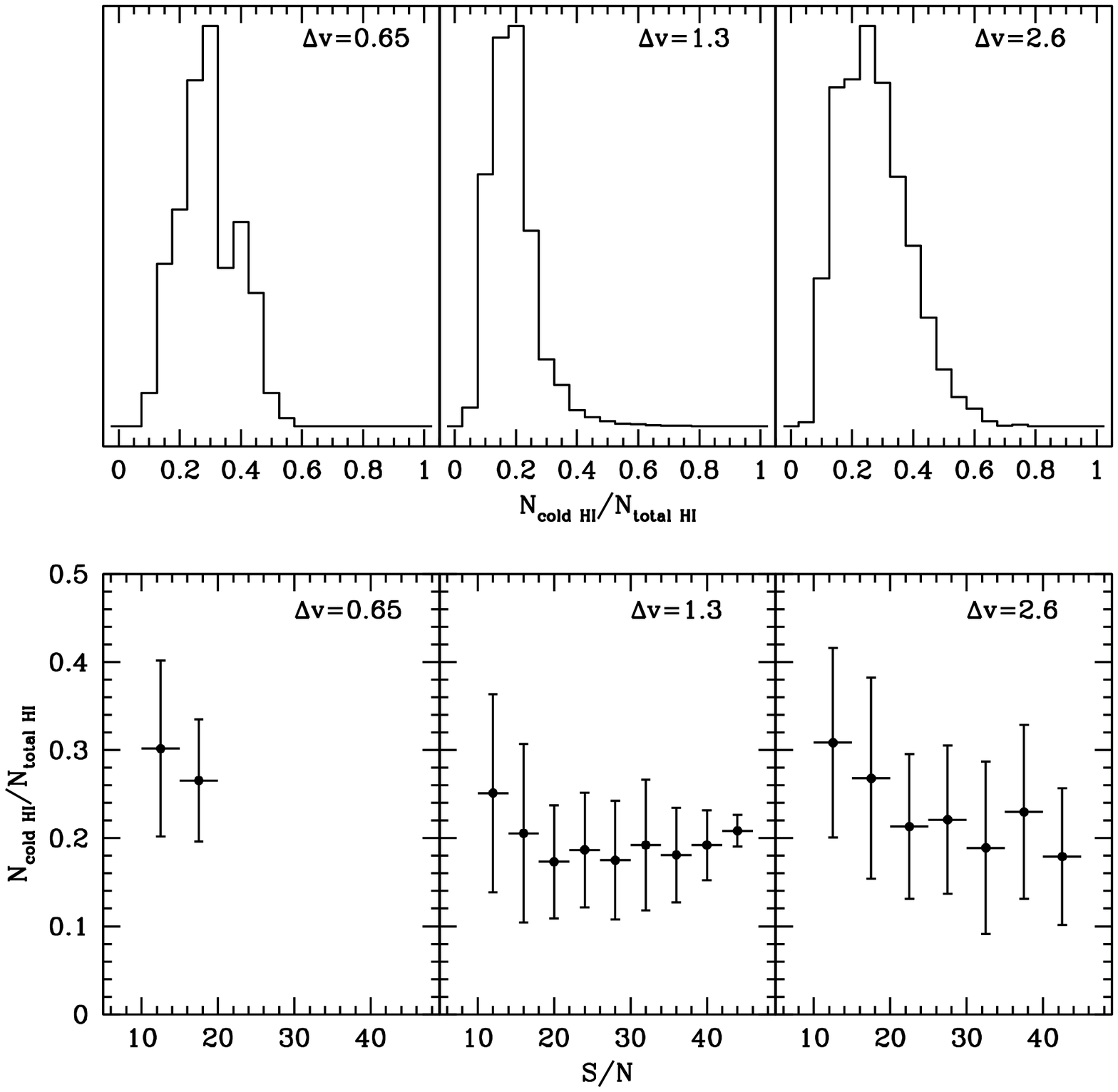}
\caption{{\it Top:} Normalized histograms showing the contribution to the total line strength of the
cold \HI\ for each velocity 
resolution.  {\it Bottom:}  The average cold-to-total flux ratio as a function of S/N for each velocity resolution.  The vertical
error bars are the dispersions in the bins.  The horizontal bar over
each point shows the bin size from which the average was computed.  For each panel we have omitted the locations 
where a single Gaussian profile with a velocity dispersion of less than 6 km s$^{-1}$ best
fit the data.  The cold \HI~typically constitutes only 20\% of the total line flux for locations which contain both cold and
warm \HI.
\label{histrat}}
\end{figure}

Another, perhaps more reasonable way to calculate an upper limit to the cold-to-total \HI~mass fraction is to find the typical 
fractional flux contributed to the cold component along each line-of-sight and assume this fraction holds throughout the sample.
For locations where we find both warm and cold \HI~components, the cold \HI~typically is not the dominant phase in the ISM.  
The top row of Figure \ref{histrat} shows a normalized histogram of the flux ratio of cold and total \HI~defined as the ratio of 
the column density of the narrow Gaussian profile and the total column density along the line-of-sight.  The cold \HI~constitutes 
$\sim$20\% of the total line flux for any given line-of-sight.  The bottom panels of Figure \ref{histrat} show this same ratio as
a function of S/N for each velocity resolution.  If we assume that {\it every} line-of-sight contains cold \HI, then an upper
limit to the cold-to-total mass ratio for each galaxy is $\sim$20\%.  This value is what is reported by \citet{youn96,youn97} and
\citet{debl06}.  However, this value is also likely too high since it requires the entire galaxy to contain cold \HI~gas.
\citet{dick00} observed the Small Magellanic Cloud in 21 cm absorption and found that the mass fraction of cold \HI~to be less
than 15\% of the total \HI~content.  It seems likely that the true value of the cold \HI~mass fractions of the low mass 
galaxies in our sample are between our lower limits and $\sim$15\%.

\citet{begu10} observed 21-cm Milky Way absorption towards 12 background radio continuum sources.  They were able to determine the flux contribution
of the cold \HI~along many of these sight lines.  They find that the cold \HI~contributes from a low of $\sim$2\% to as high as $\sim$69\%,
of the total flux determined in the same manner, with a median of $\sim$30\%.  While a direct comparison between the optically thick cold
\HI~observed by \citet{begu10} and our optically thin emission is not obvious, we would not expect the fraction of cold \HI~in emission 
to be drastically different.  Our result of $\sim$20\% is slightly lower than their median value, but does
not appear to be at odds with their result.

\subsection{Cold \HI~Locations that Lack a Warm Component \label{nowarm}}

\begin{figure}
\includegraphics[width=175mm]{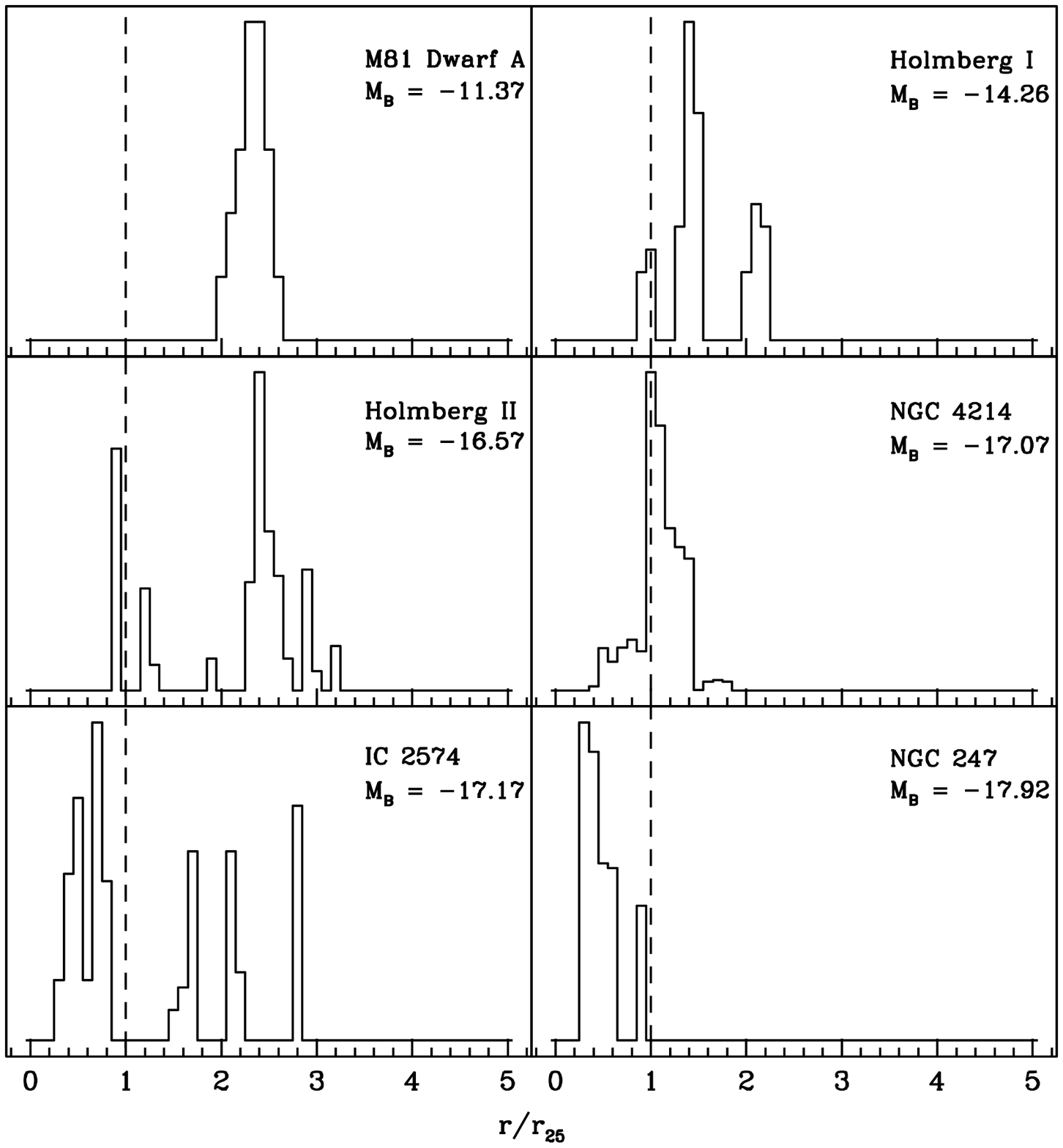}
\caption{Radial distribution plots of the locations of the cold \HI~that lack a warm ($\sigma>6$ km s$^{-1}$) component as a function of radius.  
The x-axis has been normalized by the 25 mag arcsec$^{-2}$ 
radius (vertical dashed line).  The galaxies have been ordered faint (M81 dwarf A) to bright 
(NGC 247) absolute $B$-band magnitude.  A significant fraction of the cold \HI~is outside of the optical radius of each galaxy, except for NGC 247.
\label{muchoblooshto}}
\end{figure}

As discussed in \S\ref{methods}, some lines-of-sight were best fit by a single Gaussian profile with a velocity dispersion of less than 6 km
s$^{-1}$.  These best fit single Gaussian profiles were found in 6 of the 23 galaxies with cold \HI~detections.  Most of these six galaxies 
have a large fraction of their \HI~at galactocentric radii beyond the $\sim$25 mag arcsec$^{-2}$ optical
level, and the majority of these single Gaussian fits are also located at larger radii.  Figure 
\ref{muchoblooshto} shows the radial distribution of the cold \HI~locations that lack a warm component.  The galaxies have been 
ordered from low (M81~Dwarf~A) to high (NGC~247) absolute $B$-band luminosity.  Two of these galaxies, M81~Dwarf~A and Holmberg~I, have 
large central holes in their \HI~distributions (see \citealt{warr11} and references therein) where much of their gas is pushed beyond the 
optical radius.  Only NGC~247 has a majority of these locations within the optical radius.

\begin{figure}
\includegraphics[width=175mm]{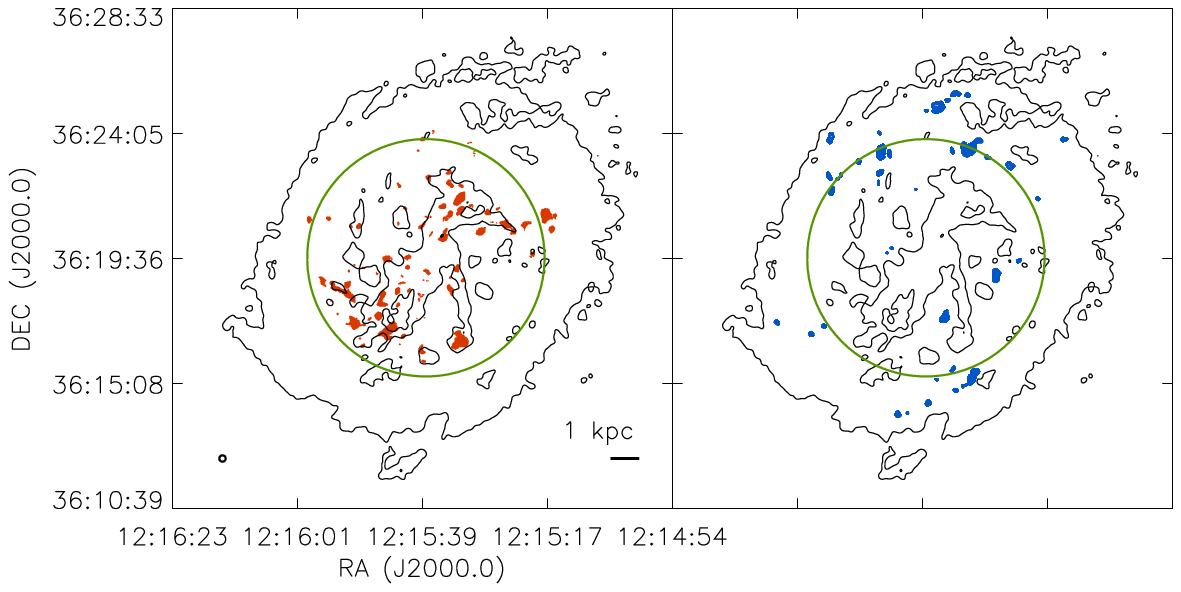}
\caption{{\it Left:} The cold \HI~distribution (red) of NGC~4214 for those locations best fit by double Gaussians profiles only.  
{\it Right:} The cold \HI~distribution (blue) for those locations best fit by single Gaussian profiles with velocity
dispersions less than 6 km s$^{-1}$ only.  The contours represent the 
10$^{20}$ and 10$^{21}$ cm$^{-2}$ total \HI~column densities.  The green circle approximates the 25 mag arcsec$^{-2}$ optical 
level.  The beam is shown at lower left.  The majority of the cold \HI~detections described by a single Gaussian profile are outside 
of the bulk of the stellar distribution. 
\label{n42142plot}}
\end{figure}

Figure \ref{n42142plot} shows an example of the difference between those areas that contain a warm component and those that lack one 
for NGC~4214.  The left panel shows only those locations that contain both a warm and
cold \HI~component while the right panel also includes the locations that lack a warm component.  The contours correspond to the 
10$^{20}$ and 10$^{21}$ cm$^{-2}$ total \HI~column densities and the blue circle approximates the 25 mag arcsec$^{-2}$ level 
(uncorrected for inclination and position angle).  It is immediately apparent that the majority of the best fit single 
Gaussian profiles fall well beyond the bulk of the stellar population and far from any heat sources.  \citet{tamb09} provides evidence
that the warm \HI~velocity dispersions inside the optical radius of disk galaxies is driven by stellar processes, while areas beyond 
the optical radius may be driven by thermal broadening and/or magneto-rotational instabilities.  These authors also point out that the 
velocity dispersions decline as a function of radius, falling below the typical 10 km s$^{-1}$ beyond
the optical radius.  We also see this declining trend in the velocity dispersion with radius (see Figure
\ref{disprad}).  These 
results are suggestive that cold \HI~can be the dominant constituent of the ISM only in the lower UV radiation field beyond the 
$\sim$25 mag arcsec$^{-2}$ optical level.

\subsection{Comparing the Cold \HI~Gas Mass with Molecular Gas Masses and Star Formation Rates \label{h2sfr}}

We next compare the cold \HI~gas masses to the molecular gas masses and star formation rates (SFRs) in each galaxy.  To make these comparisons
we compute the SFRs and estimate the amount of molecular hydrogen each galaxy contains.  The relationship between the SFR and molecular hydrogen is
discussed in detail in \S\ref{intro}.  We used the SFRs given by both FUV and H$\alpha$
luminosities to estimate the amount of molecular hydrogen.  

Table \ref{SFRs} lists the SFRs derived from FUV and H$\alpha$ luminosities taken from the literature for each
galaxy with a detected cold \HI~signature.  The columns are: (1) galaxy name, (2) mass of the cold \HI~($M_{coldHI}$) computed using
the lower limit mass fraction from Table \ref{indres} multiplied by the total \HI~mass given in Table \ref{galprop}, (3) apparent FUV magnitude 
(m$_{FUV}$; \citealt{lee11}), (4) the logarithm of the FUV luminosity (L$_{FUV}$), (5) the FUV SFR (SFR$_{FUV}$) computed using the relation given in 
\citet{sali07}, (6) the H$_{2}$ gas mass ($M_{H_{2}}^{FUV}$) computed from SFR$_{FUV}$ assuming a 100 Myr timescale and a star 
formation efficiency of 10\% ($M_{H_{2}}^{FUV}$ = SFR$_{FUV}$ $\times$ 100 Myr / 0.1), (7) the approximate fraction of the cold ISM attributed to the cold 
\HI~($\mathcal{M}_{FUV}$) defined as $M_{coldHI}$/($M_{coldHI}$ + $M_{H_{2}}^{FUV}$), (8) the logarithm of the H$\alpha$ luminosity (L$_{H\alpha}$; 
\citealt{kenn08}), (9) the SFR derived from L$_{H\alpha}$ using the relation in \citet{kenn98}, (10) the H$_{2}$ gas mass ($M_{H_{2}}^{H\alpha}$) 
computed from SFR$_{H\alpha}$ assuming a 5 Myr timescale and a star formation efficiency of 10\% ($M_{H_{2}}^{H\alpha}$ = SFR$_{FUV}$
$\times$ 5 Myr / 0.1), 
and (11) the approximate fraction of the cold ISM 
attributed to the cold \HI~($\mathcal{M}_{H\alpha}$) defined as $M_{coldHI}$/($M_{coldHI}$ + $M_{H_{2}}^{H\alpha}$).  Using our lower limit mass
fraction from the spectra above S/N of 10 and extrapolating it over the entire galaxy provides a lower limit to the total cold \HI~mass.

The expected relationship between the currently observed cold \HI~and the computed molecular gas masses is not immediately clear.  The mass fractions of 
the cold ISM attributed to the cold \HI~is lower for H$_{2}$ masses computed from the longer timescale SFR$_{FUV}$ (Table \ref{SFRs}, column 7) than they are for 
those computed from SFR$_{H\alpha}$ (Table \ref{SFRs}, column 11).  The time scale to
convert the cold \HI~into molecular hydrogen is likely dependent upon the metallicity of the environment (a longer time scale is expected for lower
metallicities) and local UV radiation field (a shorter time scale is expected for lower UV radiation).  Since dwarf galaxies have both low
metallicities and different ranges of UV radiation strengths, it is not obvious which molecular gas mass 
estimate provides the best comparison.  Unfortunately, our current data do not allow us to 
investigate these timescales further.  

\begin{figure}
\includegraphics[width=175mm]{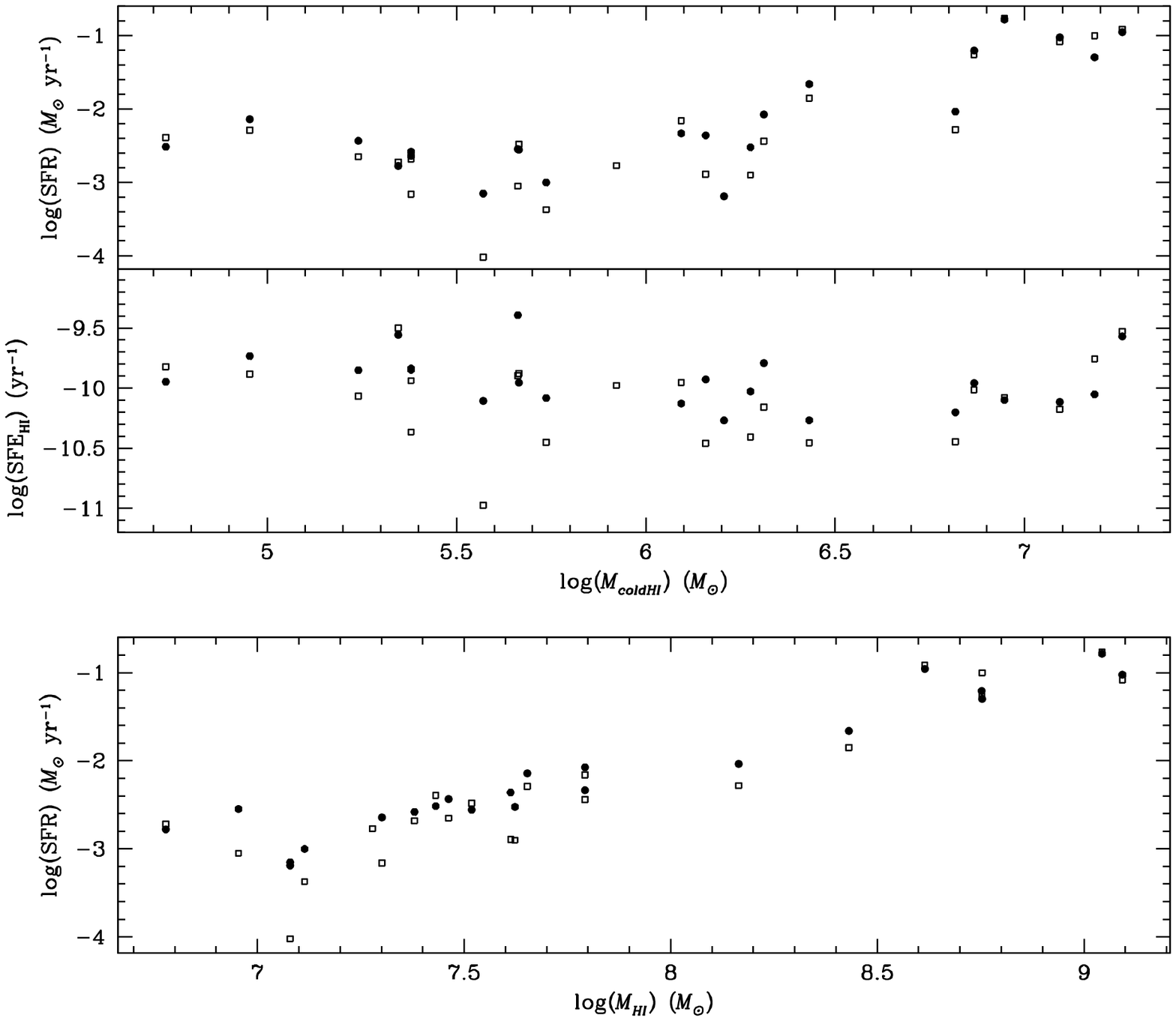}
\caption{{\it Top:} SFR$_{FUV}$ (filled circles) and SFR$_{H\alpha}$ (open squares) as a function of the cold \HI~mass.  
{\it Middle:} The SFR efficiency (SFE$_{HI}$) defined as the SFR divided by the total \HI~gas mass as a
function of the cold \HI~gas mass.  {\it Bottom:} The SFR as a function of the total \HI~gas mass.  The SFR shows a linear trend with
both cold and total \HI~mass.
\label{sfrfig}}
\end{figure}

The top and bottom panels of Figure \ref{sfrfig} show plots of the SFR$_{FUV}$ (filled circles) and SFR$_{H\alpha}$ (open squares) as a function
of the cold \HI~mass and total \HI~mass, respectively.  If the 
cold \HI~is related to the star formation in any direct way, we would expect to find a trend in this relation.  Indeed, the higher the cold \HI~gas mass, the 
higher the SFR, although this same trend is seen with the total \HI~gas mass and may just be a consequence of the more \HI~gas available, the higher the SFR.  
The middle panel shows the SFR efficiency (SFE$_{HI}$) defined as the SFR divided by the total \HI~gas mass as a
function of the cold \HI~gas mass.  The SFE$_{HI}$ is roughly constant over our range of cold \HI~masses.  The differences between the derived FUV and H$\alpha$ star 
formation rates has been previously reported in \citet{lee09} and is due at least in part to the fact that the SFRs are averaged over different timescales.  Environmental
conditions may also influence SFR$_{H\alpha}$ by leakage of the ionizing photons from the star forming regions (see \citealt{rela12} for details).

We can compare our molecular gas mass estimates derived from the SFRs to a direct observational estimate of the molecular gas in 
NGC~4214 \citep{walt01} to see if our values are reasonable.  These authors observed CO emission in the inner $\sim$1$\farcm$6$\times$1$\farcm$6 of the star forming disk of 
NGC~4214 (an area much smaller than the \HI~disk, see \S\ref{nowarm}) and derived a molecular gas mass of 
5.1$\times$10$^{6}$ $M_{\odot}$.  This value is similar to the 
molecular gas mass estimate we get from SFR$_{H\alpha}$ of 6.1$\times$10$^{6}$ $M_{\odot}$.  The molecular gas mass value computed by \citeauthor{walt01} 
was based on an assumed CO to H$_{2}$ conversion factor (the so called 
'X-factor') derived from observations of Galactic environments (i.e., high metallicity).  Recent work on the X-factor has shown that the relation changes
with decreasing mean visual extinction (e.g., \citealt{lero09, glov11, schr12}).  The lower metallicity environment of dIrr galaxies results in lower visual extinction 
values and thus a different X-factor value.  Depending upon the true mean visual extinction value of NGC~4214, the actual molecular gas mass could be higher 
by a factor of a few to an order of magnitude or more.  \citet{schr12} derive an X-factor value roughly 40 times higher than the Galactic value for NGC~4214.  
Using this value, the \citet{walt01} molecular gas mass increases to $\sim$2$\times$10$^{8}$ $M_{\odot}$, closer to our derived H$_{2}$ mass using SFR$_{FUV}$ of 
1$\times$10$^{8}$ $M_{\odot}$.  \citet{schr12} also derive a molecular gas mass of $\sim$2$\times$10$^{8}$ $M_{\odot}$ for NGC~4214.  The factor of 2 difference 
between the values most likely arise from our basic assumptions of a timescale of 100 Myr for the FUV emission and/or the star formation efficiency of 10\%.  
Unfortunately, the lack of CO detections for the few galaxies in our sample that have been observed for CO
prohibit us from repeating this exercise for other galaxies in our sample (e.g., \citealt{isra95}; \citealt{vert95}; \citealt{lero09}; \citealt{schr12}).  
Nevertheless, our estimates of molecular gas masses, although uncertain, are not unreasonable.  Direct comparisons between the CO detections in NGC~4214 and our cold
\HI~detections is the focus of future work.

\section{Conclusions \label{conclusions}}

\begin{figure}
\includegraphics[width=175mm]{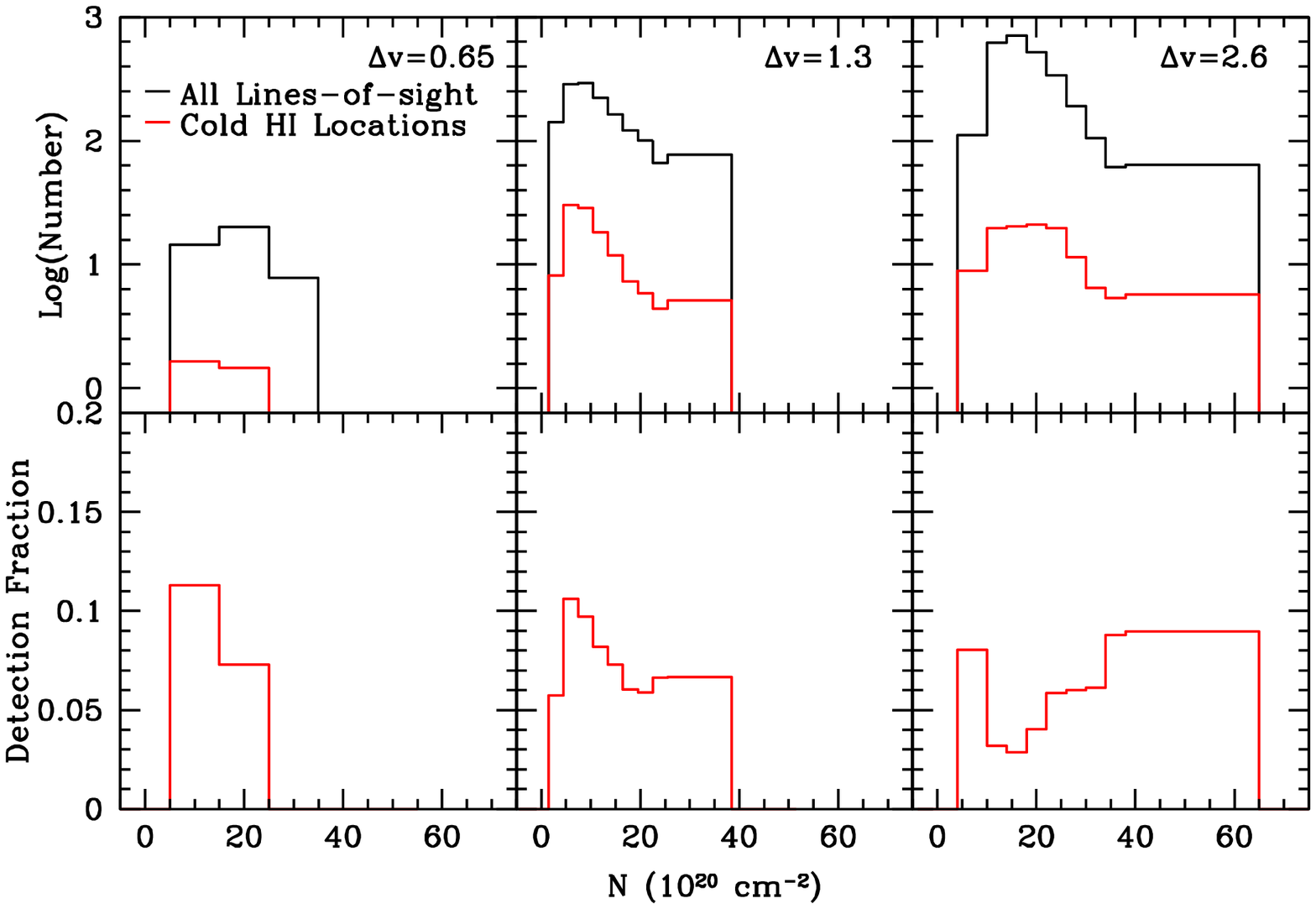}
\caption{{\it Top:} Histograms of observed column densities of our entire sample (black) and where we
find cold \HI\ for each
velocity resolution.  {\it Bottom:} The detection fraction as a function of column density.  Our
detection fractions are relatively constant across all column densities observed.
\label{cdplot}}
\end{figure}

We have observed line-of-sight \HI~emission spectra in 31 nearby, low metallicity dIrr galaxies in order to search for cold \HI~defined by
a velocity dispersion of less than 6 km s$^{-1}$.  We have detected it in 23 of 27 galaxies after quality control cuts were applied.
The cold \HI~may be the future sites of molecular cloud and star formation and to date, are the only way to potentially trace star forming 
gas at low metallicity.  Based upon our observations, we find:

\begin{itemize}
\renewcommand{\labelitemi}{$\bullet$}

\item The cold \HI~gas is found in localized regions and usually at total \HI~column densities above the canonical threshold of star formation 
of 10$^{21}$ cm$^{-2}$ \citep{skil87}. The cold \HI~in a given galaxy is also not typically associated with the very 
highest surface density gas (see Figures \ref{2plots} \& \ref{cdplot}).

\item The cold \HI~has a typical velocity dispersion of $\sim$4.5 km s$^{-1}$ (T $\lesssim$ 800 K).

\item We derive an average value to the volume filling fraction of the cold \HI~of 9\% (assuming our lower limit detections).

\item We find lower limits to the cold \HI~gas mass fractions of a few percent, consistent with some models of the multi-phase ISM \citep{mcke77}.

\item The SFE$_{HI}$ is roughly constant as a function of the total cold \HI~mass over the observed gas mass ranges.

\item The cold \HI~contributes $\sim$20\% of the line-of-sight flux in locations where we detect both a cold and warm component.

\item Cold \HI~gas that lacks a warm component is typically found at radii larger than the 25 mag arcsec$^{-2}$ optical radius.

%\item The cold \HI~is only found at total \HI~column densities above $\sim$2$\times$10$^{20}$ cm$^{-2}$ (see Figures \ref{cdplot} and 
%\ref{climit}), consistent with recent Galactic observations \citep{kane11} and independent of S/N effects.

%\item The fraction of cold \HI~detections increases with increasing N$_{\mathrm{HI}}$, indicating that these galaxies are more efficient 
%at creating cold \HI~at large values of N$_{\mathrm{HI}}$.
\end{itemize}

Future work will investigate the relationship between the cold \HI, local ISM conditions, and tracers of current star formation.  

\acknowledgements{We thank the anonymous referee for their many thoughtful comments and insightful suggestions which greatly 
improved the quality of this work.  Support for this work was provided by the
National Science Foundation collaborative research grant `Star Formation, 
Feedback, and the ISM: Time Resolved Constraints from a Large VLA Survey of Nearby Galaxies,'
grant number AST-0807710.  This material is based upon work supported by the National Science Foundation 
under Grant No. DGE-0718124.  SRW is grateful for support
from a Penrose Fellowship, a University of Minnesota Degree Dissertation Fellowship, and a NRAO 
Research Fellowship number 807515.  NRAO is operated by Associated Universities,
Inc., under cooperative agreement with the National Science Foundation.  This research
has made use of NASA's Astrophysics Data System Bibliographic
Services and the NASA/IPAC Extragalactic
Database (NED), which is operated by the Jet Propulsion
Laboratory, California Institute of Technology, under 
contract with the National Aeronautics and Space Administration.}

\clearpage

\begin{table}
\begin{center}
\tiny
\caption{Observed Galaxy Properties \label{galprop}}
\begin{tabular}{lcccccccccc}
\hline\hline
~~~~1 & 2 & 3 & 4 & 5 & 6 & 7 & 8 & 9 & 10 & 11\\
Galaxy & RA & DEC & D$^{a}$ & $M_{\mathrm{HI}}$ & M$_{B}$ & $a^{b}$ & i$^{b}$ & Beam Size & $\Delta$ v & rms$_{noise}$ \\
& (J2000.0) & (J2000.0) & (Mpc) & (10$^{7}$ $M_{\odot}$) & & (\arcmin) & (deg) &(\arcsec) & (km s$^{-1}$) & (mJy beam$^{-1}$) \\
\hline
DDO~53       & $08^{h}34^{m}07\fs2$ & $+66{\degr}10{\arcmin}54{\arcsec}$ & 3.61      & 6.2   & -13.45 & 1.6  & 30 & 11.43 & 2.6  & 0.8 \\
DDO~82       & $10^{h}30^{m}35\fs0$ & $+70{\degr}37{\arcmin}10{\arcsec}$ & 3.80      & 0.3   & -14.44 & 3.4  & 55 & 10.86 & 1.3  & 1.7 \\
DDO~99       & $11^{h}50^{m}53\fs0$ & $+38{\degr}52{\arcmin}50{\arcsec}$ & 2.59      & 4.7   & -13.37 & 4.1  & 71 & 15.93 & 1.3  & 2.0 \\
DDO~125      & $12^{h}27^{m}41\fs8$ & $+43{\degr}29{\arcmin}38{\arcsec}$ & 2.58      & 2.9   & -14.04 & 4.3  & 58 & 15.99 & 0.65 & 3.0 \\
DDO~181      & $13^{h}39^{m}53\fs8$ & $+40{\degr}44{\arcmin}21{\arcsec}$ & 3.14      & 2.4   & -12.94 & 2.3  & 57 & 13.14 & 1.3  & 1.6 \\
DDO~183      & $13^{h}50^{m}51\fs1$ & $+38{\degr}01{\arcmin}16{\arcsec}$ & 3.22      & 2.0   & -13.08 & 2.2  & 75 & 12.81 & 1.3  & 1.5 \\
DDO~187      & $14^{h}15^{m}56\fs5$ & $+23{\degr}03{\arcmin}19{\arcsec}$ & 2.21      & 1.2   & -12.43 & 1.7  & 42 & 18.67 & 1.3  & 2.1 \\
DDO~190      & $14^{h}24^{m}43\fs5$ & $+44{\degr}31{\arcmin}33{\arcsec}$ & 2.79      & 4.1   & -14.14 & 1.8  & 28 & 14.78 & 2.6  & 0.7 \\
GR8          & $12^{h}58^{m}40\fs4$ & $+14{\degr}13{\arcmin}03{\arcsec}$ & 2.08      & 0.6   & -12.00 & 1.1  & 25 & 19.83 & 0.65 & 4.4 \\
Holmberg~I   & $09^{h}40^{m}32\fs3$ & $+71{\degr}10{\arcmin}56{\arcsec}$ & 3.90      & 14.6  & -14.26 & 3.6  & 37 & 10.58 & 2.6  & 1.2 \\
Holmberg~II  & $08^{h}19^{m}05\fs0$ & $+70{\degr}43{\arcmin}12{\arcsec}$ & 3.38      & 56.6  & -16.57 & 7.9  & 31 & 12.20 & 2.6  & 1.5 \\
IC~2574      & $10^{h}28^{m}27\fs7$ & $+68{\degr}24{\arcmin}59{\arcsec}$ & 3.80      & 123.9 & -17.17 & 13.2 & 68 & 10.86 & 2.6  & 0.9 \\
KDG~73       & $10^{h}52^{m}55\fs3$ & $+69{\degr}32{\arcmin}45{\arcsec}$ & 4.03      & 0.05  & -10.75 & 0.6  & 35 & 10.24 & 0.65 & 2.0 \\
KK~230       & $14^{h}07^{m}10\fs7$ & $+35{\degr}03{\arcmin}37{\arcsec}$ & 1.97      & 0.07  & -8.49  & 0.6  & 35 & 20.94 & 0.65 & 3.8 \\
KKH~98       & $23^{h}45^{m}34\fs0$ & $+38{\degr}43{\arcmin}04{\arcsec}$ & 2.54      & 0.3   & -10.29 & 1.1  & 58 & 16.24 & 0.65 & 2.8 \\
M81~Dwarf~A  & $08^{h}23^{m}55\fs1$ & $+71{\degr}01{\arcmin}56{\arcsec}$ & 3.44      & 1.2   & -11.37 & 1.3  & 24 & 11.99 & 1.3  & 0.9 \\
M81~Dwarf~B  & $10^{h}05^{m}30\fs6$ & $+70{\degr}21{\arcmin}52{\arcsec}$ & 5.3$^{b}$ & 2.7   & -14.23 & 0.9  & 49 & 7.78  & 2.6  & 0.6 \\
MCG09-20-131 & $12^{h}15^{m}46\fs7$ & $+52{\degr}23{\arcmin}15{\arcsec}$ & 1.6$^{c}$ & 1.2   & -12.36 & 1.2  & 77 & 25.78 & 1.3  & 3.3 \\
NGC~247      & $00^{h}47^{m}08\fs3$ & $-20{\degr}45{\arcmin}36{\arcsec}$ & 3.52      & 110.6 & -17.92 & 21.4 & 72 & 11.72 & 2.6  & 1.3 \\
NGC~2366     & $07^{h}28^{m}53\fs4$ & $+69{\degr}12{\arcmin}51{\arcsec}$ & 3.21      & 56.7  & -15.85 & 7.3  & 72 & 12.85 & 2.6  & 1.0 \\
NGC~3109     & $10^{h}03^{m}07\fs2$ & $-26{\degr}09{\arcmin}36{\arcsec}$ & 1.27      & 27.0  & -15.18 & 19.7 & 86 & 32.48 & 1.3  & 8.5 \\
NGC~3741     & $11^{h}36^{m}06\fs4$ & $+45{\degr}17{\arcmin}07{\arcsec}$ & 3.24      & 8.1   & -13.01 & 2.0  & 58 & 12.73 & 1.3  & 2.1 \\
NGC~4163     & $12^{h}12^{m}08\fs9$ & $+36{\degr}10{\arcmin}10{\arcsec}$ & 2.87      & 0.9   & -13.76 & 1.9  & 34 & 14.37 & 0.65 & 2.2 \\
NGC~4190     & $12^{h}13^{m}44\fs6$ & $+36{\degr}38{\arcmin}00{\arcsec}$ & 3.5$^{b}$ & 4.5   & -14.20 & 1.7  & 29 & 11.79 & 1.3  & 1.5 \\
NGC~4214     & $12^{h}15^{m}39\fs2$ & $+36{\degr}19{\arcmin}37{\arcsec}$ & 3.04      & 41.2  & -17.07 & 8.5  & 40 & 13.57 & 1.3  & 1.1 \\
Sextans~A    & $10^{h}11^{m}00\fs8$ & $-04{\degr}41{\arcmin}34{\arcsec}$ & 1.38      & 6.2   & -13.71 & 5.9  & 35 & 29.89 & 1.3  & 4.5 \\
Sextans~B    & $10^{h}00^{m}00\fs1$ & $+05{\degr}19{\arcmin}56{\arcsec}$ & 1.39      & 4.2   & -13.88 & 5.1  & 48 & 29.68 & 1.3  & 2.0 \\
UGCA~292     & $12^{h}38^{m}40\fs0$ & $+32{\degr}46{\arcmin}00{\arcsec}$ & 3.62      & 4.0   & -11.36 & 1.0  & 47 & 11.40 & 0.65 & 2.1 \\
UGC~4483     & $08^{h}37^{m}03\fs0$ & $+69{\degr}32{\arcmin}45{\arcsec}$ & 3.41      & 3.3   & -12.58 & 1.2  & 56 & 12.10 & 2.6  & 0.8 \\
UGC~8508     & $13^{h}30^{m}44\fs4$ & $+54{\degr}54{\arcmin}36{\arcsec}$ & 2.58      & 1.9   & -12.95 & 1.7  & 55 & 15.99 & 0.65 & 2.3 \\
UGC~8833     & $13^{h}54^{m}48\fs7$ & $+35{\degr}50{\arcmin}15{\arcsec}$ & 3.08      & 1.3   & -12.31 & 0.9  & 28 & 13.39 & 2.6  & 0.6 \\
\hline
\end{tabular}
\end{center}
$^{a}$\citet{dalc09}; $^{b}$\citet{kara04}; $^{c}$MCG09-20-131 may have a greater distance than indicated due to ambiguities in measuring
the TRGB.
\end{table}

\clearpage

\begin{table}
\begin{center}
\caption{Simulated Spectra Results \label{simtab}}
\begin{tabular}{lcccccc}
\hline\hline
~~~~1 & 2 & 3 & 4 & 5 & 6 & 7 \\
Velocity   & A$_{b,sim}$/A$_{b,extr}$ & $\Delta$v$_{broad}$ & $\Delta$$\sigma$$_{broad}$ & A$_{n,sim}$/A$_{n,extr}$ & $\Delta$v$_{nar}$ & $\Delta$$\sigma$$_{nar}$ \\
Resolution &                          &  (km/s)             &  (km/s)                    &                          &  (km/s)           &  (km/s)                  \\
\hline
0.65 & 1.0$\pm$0.1 &  0.0$\pm$0.3 & 0.0$\pm$0.6 & 1.0$\pm$0.1 & 0.0$\pm$0.2 & 0.0$\pm$0.3 \\
1.30 & 1.0$\pm$0.2 &  0.0$\pm$0.5 & 0.0$\pm$0.7 & 1.0$\pm$0.1 & 0.1$\pm$0.2 & 0.1$\pm$0.3 \\
2.60 & 1.0$\pm$0.2 & -0.2$\pm$0.8 & 0.0$\pm$0.9 & 1.0$\pm$0.1 & 0.5$\pm$0.4 & 0.1$\pm$0.5 \\
\hline
\end{tabular}
\end{center}
\end{table}

\clearpage

\begin{table}
\begin{center}
\caption{Individual Galaxy Properties \label{fitres}}
\begin{tabular}{lccccccccccc}
\hline\hline
~~~~1 & 2 & 3 & 4 & 5 & 6 \\
Galaxy & N$_{t}$ & $<$S/N$>_t$ & N$_{\mathrm{HI},min}$ & N$_{\mathrm{HI},peak}$ & $<$N$_{\mathrm{HI}}$$>$ \\
& & & (10$^{19}$ cm$^{-2}$) & (10$^{21}$ cm$^{-2}$) & (10$^{20}$ cm$^{-2}$) \\
\hline
DDO~53       & 48   & 14.8    & 6.76  & 3.19    & 16.0    \\
DDO~82       & 0    & \nodata & 19.20 & \nodata & \nodata \\
DDO~99       & 19   & 12.7    & 14.28 & 2.38    & 13.0    \\
DDO~125      & 6    & 10.8    & 14.07 & 1.62    & 11.3    \\
DDO~181      & 9    & 11.2    & 13.69 & 1.53    & 11.3    \\
DDO~183      & 11   & 11.9    & 11.78 & 1.97    & 14.1    \\
DDO~187      & 13   & 14.8    & 7.78  & 2.59    & 14.0    \\
DDO~190      & 46   & 20.6    & 3.86  & 3.55    & 15.8    \\
GR8          & 3    & 10.7    & 12.97 & 1.06    & 9.23    \\
Holmberg~I   & 24   & 11.1    & 11.84 & 2.47    & 16.6    \\
Holmberg~II  & 207  & 12.0    & 11.11 & 3.82    & 18.1    \\
IC~2574      & 1026 & 13.1    & 8.43  & 4.83    & 15.8    \\
KDG~73       & 0    & \nodata & 27.69 & \nodata & \nodata \\
KK~230       & 0    & \nodata & 10.77 & \nodata & \nodata \\
KKH~98       & 0    & \nodata & 15.55 & \nodata & \nodata \\
M81~Dwarf A  & 3    & 10.7    & 6.91  & 0.64    & 5.67    \\
M81~Dwarf B  & 9    & 11.8    & 10.93 & 3.53    & 22.9    \\
MCG09-20-131 & 3    & 14.0    & 7.16  & 1.97    & 12.7    \\
NGC~247      & 848  & 12.9    & 15.10 & 5.24    & 22.6    \\
NGC~2366     & 511  & 16.2    & 6.54  & 6.83    & 21.0    \\
NGC~3109     & 404  & 17.6    & 10.49 & 3.83    & 13.0    \\
NGC~3741     & 9    & 11.0    & 17.62 & 3.07    & 22.0    \\
NGC~4163     & 4    & 11.1    & 13.89 & 1.48    & 11.1    \\
NGC~4190     & 20   & 12.0    & 14.63 & 3.32    & 22.6    \\
NGC~4214     & 488  & 12.6    & 6.60  & 3.26    & 10.5    \\
Sextans~A    & 133  & 16.5    & 6.53  & 4.02    & 10.2    \\
Sextans~B    & 173  & 17.0    & 2.65  & 1.41    & 4.40    \\
UGCA~292     & 20   & 12.0    & 24.97 & 3.55    & 23.6    \\
UGC~4483     & 28   & 14.9    & 7.53  & 2.92    & 15.1    \\
UGC~8508     & 12   & 13.1    & 11.42 & 2.47    & 16.8    \\
UGC~8833     & 16   & 16.6    & 4.38  & 2.21    & 12.0    \\
\hline
\end{tabular}
\end{center}
\end{table}

\clearpage

\begin{table}
\begin{center}
\small
\caption{Individual Galaxy Results\label{indres}}
\begin{tabular}{lccccccc}
\hline\hline
~~~~1 & 2 & 3 & 4 & 5 & 6 & 7 & 8 \\
Galaxy & $<$S/N$>_c$ & $\mathcal{F}_{fill}$ & $\mathcal{F}_{mass}^{low}$ & $\mathcal{F}_{mass}^{up}$ & $<$$\sigma_{n}$$>$
& $<$$\sigma_{b}$$>$ &$<$$\sigma_{s}$$>$ \\
& & (\%) & (\%) & (\%) & (km s$^{-1}$) & (km s$^{-1}$) & (km s$^{-1}$) \\
\hline
DDO~53       & 18.6    & 6.1     & 2.0     & 15.9 & 4.36    & 11.92   & 9.83    \\
%DDO~82       & \nodata & \nodata & \nodata & \nodata & \nodata & \nodata \\
DDO~99       & \nodata & \nodata & \nodata & 38.3 & \nodata & \nodata & 8.94    \\
DDO~125      & 11.2    & 2.0     & 0.6     & 43.3 & 4.34    & 10.75   & 8.38    \\
DDO~181      & 13.6    & 3.4     & 1.0     & 42.1 & 3.74    & 11.89   & 8.43    \\
DDO~183      & 12.7    & 7.0     & 1.2     & 28.3 & 3.12    & 12.11   & 10.02   \\
DDO~187      & 17.1    & 21.8    & 3.1     & 18.2 & 3.79    & 14.44   & 13.31   \\
DDO~190      & 28.1    & 11.9    & 3.5     & 11.1 & 4.74    & 12.17   & 9.95    \\
GR8          & 10.4    & 10.7    & 3.7     & 47.5 & 5.30    & 12.12   & 9.62    \\
Holmberg~I   & 11.8    & 5.9     & 4.5     & 30.0 & 5.20    & 12.60   & 8.27    \\
Holmberg~II  & 12.1    & 3.3     & 1.3     & 22.2 & 5.19    & 16.00   & 10.02   \\
IC~2574      & 14.5    & 2.2     & 1.0     & 19.2 & 4.97    & 13.12   & 9.47    \\
%KDG~73       & \nodata & \nodata & \nodata & \nodata & \nodata & \nodata \\
%KK~230       & \nodata & \nodata & \nodata & \nodata & \nodata & \nodata \\
%KKH~98       & \nodata & \nodata & \nodata & \nodata & \nodata & \nodata \\
M81~Dwarf~A  & 10.5    & 13.1    & 13.4    & 50.2 & 5.86    & \nodata & 6.88    \\
M81~Dwarf~B  & 11.1    & 0.8     & 0.2     & 25.0 & 3.77    & 12.15   & 12.14   \\
MCG09-20-131 & \nodata & \nodata & \nodata & 3.6  & \nodata & \nodata & 11.99   \\
NGC~247      & 12.5    & 2.2     & 0.8     & 23.7 & 5.15    & 17.10   & 10.81   \\
NGC~2366     & 18.5    & 11.8    & 2.7     & 12.5 & 5.02    & 17.31   & 14.42   \\
NGC~3109     & 23.6    & 3.7     & 1.0     & 28.2 & 4.69    & 11.88   & 9.85    \\
NGC~3741     & \nodata & \nodata & \nodata & 27.9 & \nodata & \nodata & 10.61   \\
NGC~4163     & 11.4    & 15.7    & 5.1     & 41.8 & 5.12    & 11.27   & 9.27    \\
NGC~4190     & 12.8    & 0.8     & 0.2     & 21.9 & 5.60    & 13.91   & 12.30   \\
NGC~4214     & 13.0    & 13.0    & 4.4     & 23.5 & 4.77    & 14.90   & 10.72   \\
Sextans~A    & 25.0    & 10.7    & 3.3     & 23.2 & 4.60    & 12.71   & 11.52   \\
Sextans~B    & 25.1    & 13.7    & 4.5     & 22.8 & 4.10    & 11.13   & 8.89    \\
UGCA~292     & \nodata & \nodata & \nodata & 36.8 & \nodata & \nodata & 10.11   \\
UGC~4483     & 20.7    & 3.8     & 1.4     & 18.2 & 4.77    & 14.30   & 10.43   \\
UGC~8508     & 14.6    & 18.9    & 4.4     & 23.7 & 4.87    & 12.74   & 11.75   \\
UGC~8833     & 21.6    & 12.6    & 4.2     & 14.2 & 4.39    & 11.75   & 9.12    \\
\hline
\end{tabular}
\end{center}
\end{table}

\clearpage

\begin{table}
\begin{center}
\tiny
\caption{Star Formation Rates and Molecular Gas Masses\label{SFRs}}
\begin{tabular}{lcccccccccc}
\hline\hline
~~~~1  & 2                         & 3             & 4              & 5                          & 6                      & 7                    & 8                        & 9                           & 10                     & 11                        \\
Galaxy & $M_{coldHI}$              & m$_{FUV}^{a}$ & log(L$_{FUV}$) & SFR$_{FUV}$                & $M_{H_{2}}^{FUV}$      & $\mathcal{M}_{FUV}$  & log(L$_{H\alpha}$)$^{b}$ & SFR$_{H\alpha}$             & $M_{H_{2}}^{H\alpha}$  & $\mathcal{M}_{H\alpha}$   \\
       & (10$^{5}$ $M_{\odot}$)    & (mag)         & (erg/s/Hz)     & (10$^{-3}$ $M_{\odot}$/yr) & (10$^{6}$ $M_{\odot}$) &                      & (erg/s)                  & (10$^{-3}$ $M_{\odot}$/yr)  & (10$^{5}$ $M_{\odot}$) &                           \\
\hline
DDO 53       & 12.4  & 15.32   & 25.63   & 4.60    & 4.60    & 0.21    & 38.94   & 6.88    & 3.44    & 0.78    \\
DDO 125      & 1.74  & 14.84   & 25.53   & 3.70    & 3.70    & 0.05    & 38.45   & 2.23    & 1.11    & 0.61    \\
DDO 181      & 2.40  & 15.63   & 25.39   & 2.62    & 2.62    & 0.08    & 38.42   & 2.08    & 1.04    & 0.70    \\
DDO 183      & 2.40  & 15.84   & 25.32   & 2.28    & 2.28    & 0.10    & 37.94   & 0.69    & 0.34    & 0.87    \\
DDO 187      & 3.72  & 16.3    & 24.81   & 0.70    & 0.70    & 0.35    & 37.08   & 0.10    & 0.05    & 0.99    \\
DDO 190      & 14.4  & 14.82   & 25.61   & 4.36    & 4.36    & 0.25    & 38.21   & 1.28    & 0.64    & 0.96    \\
GR8          & 2.22  & 15.23   & 25.19   & 1.66    & 1.66    & 0.12    & 38.38   & 1.90    & 0.95    & 0.70    \\
Holmberg I   & 65.7  & 14.74   & 25.93   & 9.18    & 9.18    & 0.42    & 38.82   & 5.22    & 2.61    & 0.96    \\
Holmberg II  & 73.6  & 12.35   & 26.76   & 62.3    & 62.3    & 0.11    & 39.84   & 54.7    & 27.1    & 0.73    \\
IC 2574      & 123.9 & 12.15   & 26.94   & 94.8    & 94.8    & 0.12    & 40.02   & 82.7    & 41.4    & 0.75    \\
M81 Dwarf A  & 16.1  & 17.35   & 24.78   & 0.65    & 0.65    & 0.71    & \nodata & \nodata & \nodata & \nodata \\
M81 Dwarf B  & 0.54  & 16.6    & 25.45   & 3.05    & 3.05    & 0.02    & 38.71   & 4.05    & 2.03    & 0.21    \\
NGC 247      & 88.5  & 11.38   & 27.18   & 164.6   & 164.6   & 0.05    & 40.34   & 172.8   & 86.4    & 0.51    \\
NGC 2366     & 153.1 & 12.47   & 26.67   & 50.3    & 50.3    & 0.23    & 40.1    & 99.5    & 49.7    & 0.75    \\
NGC 3109     & 27.0  & 11.37   & 26.30   & 21.8    & 21.8    & 0.11    & 39.25   & 14.0    & 7.02    & 0.79    \\
NGC 4163     & 4.59  & 15.35   & 25.42   & 2.84    & 2.84    & 0.14    & 38.05   & 0.89    & 0.44    & 0.91    \\
NGC 4190     & 0.90  & 14.77   & 25.82   & 7.20    & 7.20    & 0.01    & 38.81   & 5.10    & 2.55    & 0.26    \\
NGC 4214     & 181.3 & 11.5    & 27.01   & 110.7   & 110.7   & 0.14    & 40.19   & 122.4   & 61.2    & 0.75    \\
Sextans A    & 20.5  & 12.58   & 25.89   & 8.41    & 8.41    & 0.20    & 38.66   & 3.61    & 1.81    & 0.92    \\
Sextans B    & 18.9  & 13.72   & 25.44   & 2.99    & 2.99    & 0.39    & 38.2    & 1.25    & 0.63    & 0.97    \\
UGC 4483     & 4.62  & 15.75   & 25.41   & 2.77    & 2.77    & 0.14    & 38.62   & 3.29    & 1.65    & 0.74    \\
UGC 8508     & 8.36  & \nodata & \nodata & \nodata & \nodata & \nodata & 38.33   & 1.69    & 0.85    & 0.91    \\
UGC 8833     & 5.46  & 16.64   & 24.96   & 0.99    & 0.99    & 0.35    & 37.73   & 0.42    & 0.21    & 0.96    \\
\hline
\end{tabular}
\end{center}
$^{a}$\citet{lee11}; $^{b}$\citet{kenn08}
\end{table}

\clearpage

\clearpage

\clearpage

\clearpage

\begin{figure}
\includegraphics[trim=0cm 0cm 0cm 5.5cm, clip=true, totalheight=0.25\textheight, angle=90]{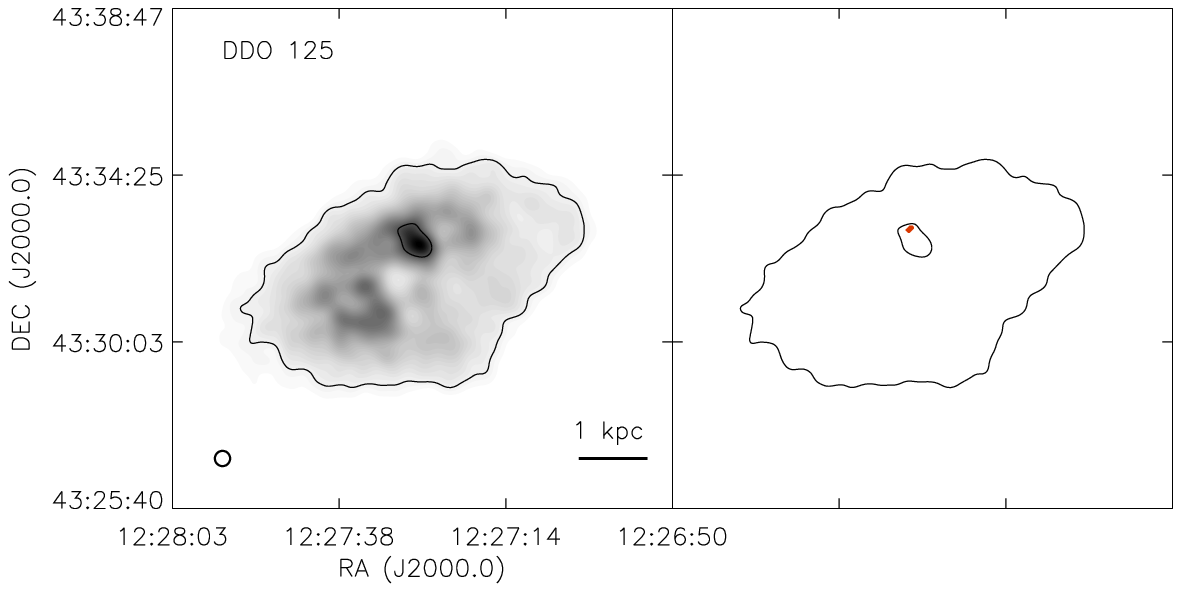} 
\includegraphics[trim=0cm 0cm 0cm 5.5cm, clip=true, totalheight=0.25\textheight, angle=90]{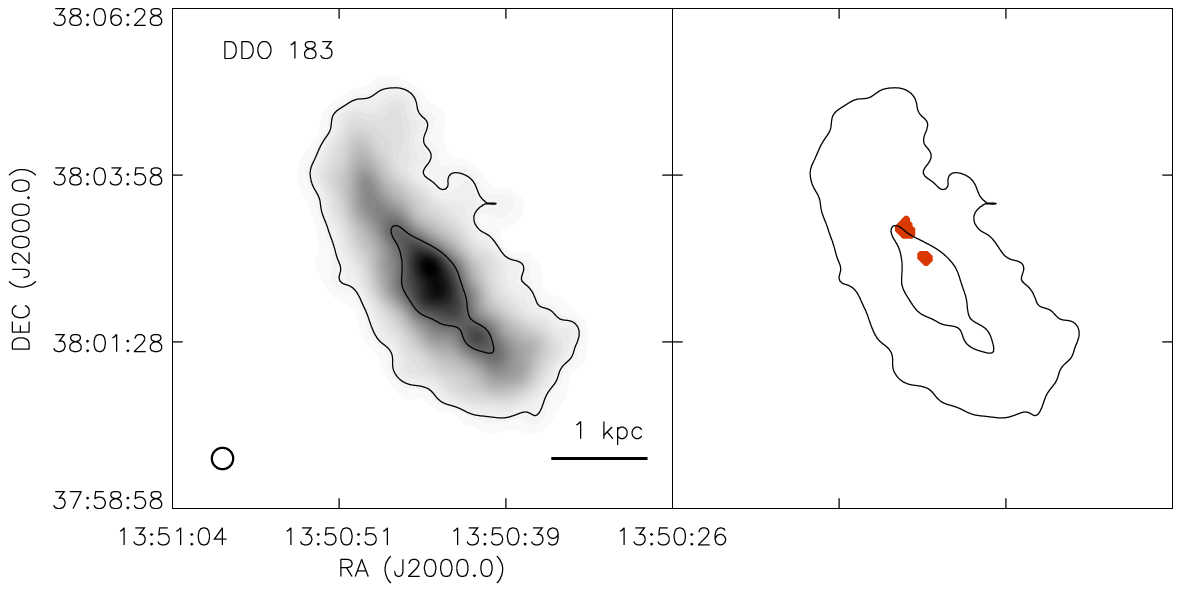}  
\includegraphics[trim=0cm 0cm 0cm 5.5cm, clip=true, totalheight=0.25\textheight, angle=90]{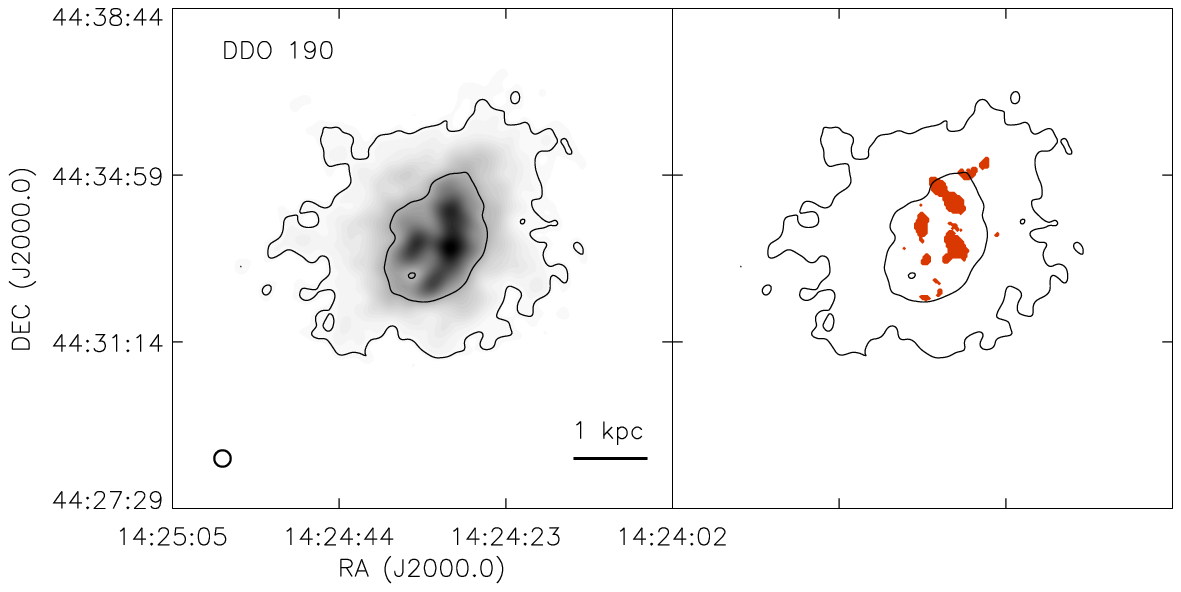} \\
\includegraphics[trim=0cm 0cm 0cm 5.5cm, clip=true, totalheight=0.25\textheight, angle=90]{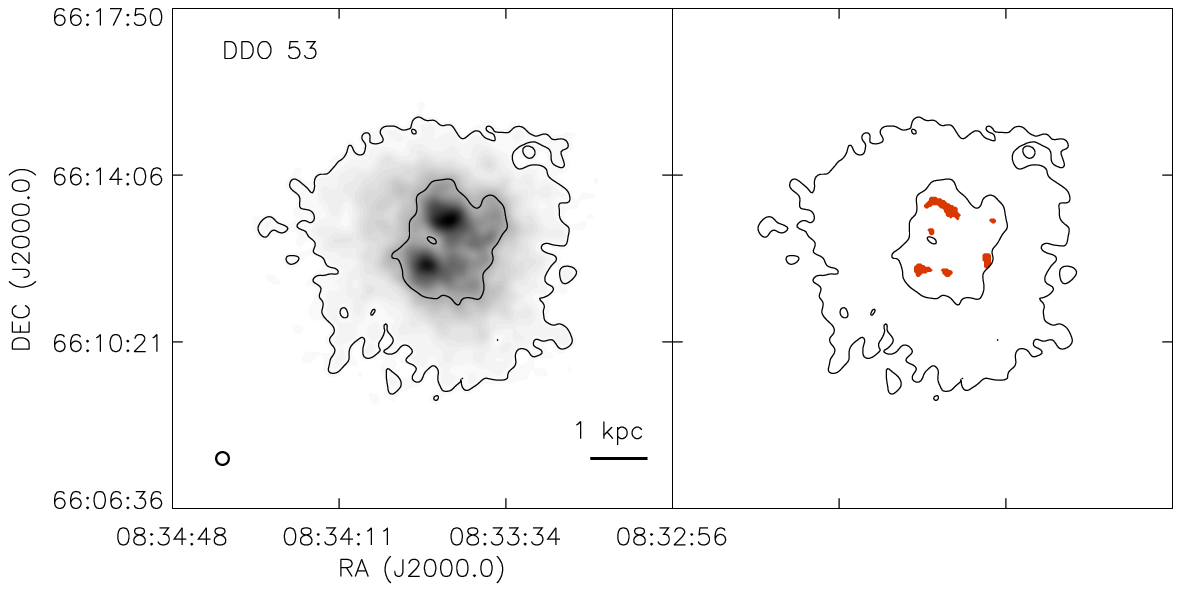} 
\includegraphics[trim=0cm 0cm 0cm 5.5cm, clip=true, totalheight=0.25\textheight, angle=90]{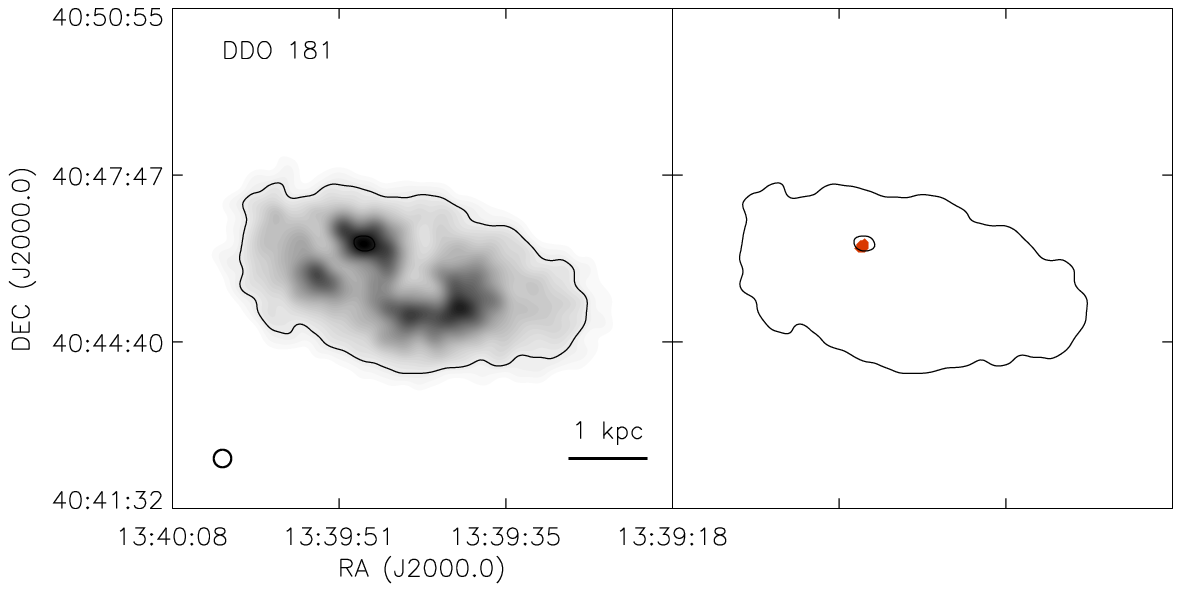}  
\includegraphics[trim=0cm 0cm 0cm 5.5cm, clip=true, totalheight=0.25\textheight, angle=90]{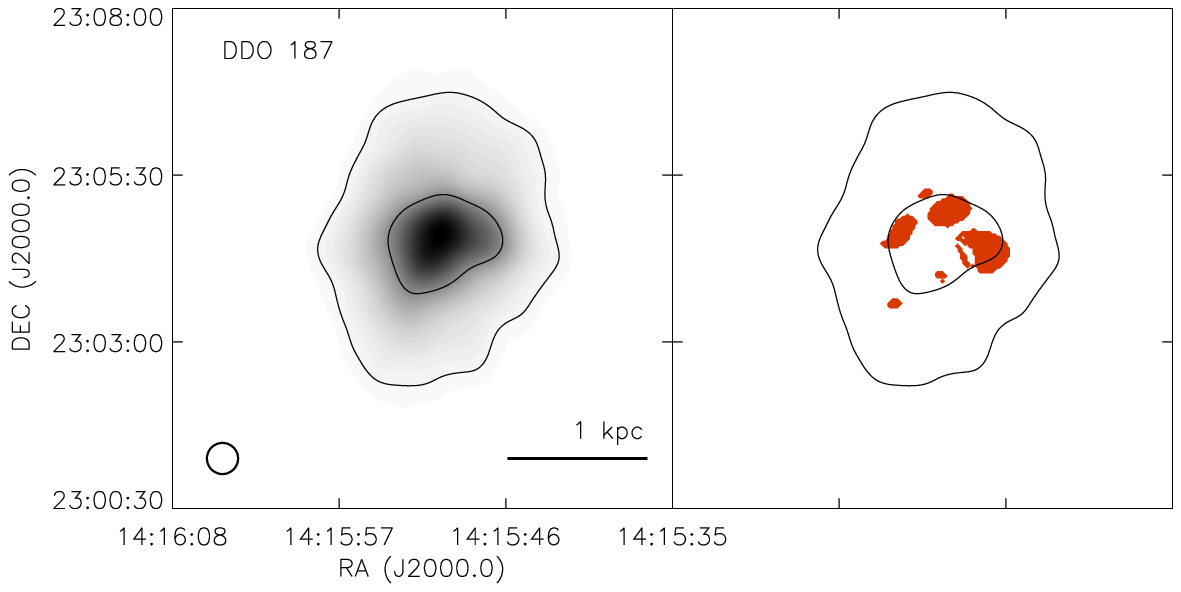}
\caption{{\it Left:} The total \HI~integrated intensity map. {\it Right:} The cold \HI~locations best described by a double
Gaussian (red) and single Gaussian (blue).  The contours for all panels represent the
$10^{20}$ and $10^{21}$ cm$^{-2}$ column density levels.  The scale bar at the bottom right denotes a linear size of 1 kpc.  
The 200 pc beam is shown at the lower left.  \label{2plots}}
\end{figure}

\clearpage

\begin{figure}
\figurenum{\ref{2plots}}
\includegraphics[trim=0cm 0cm 0cm 5.5cm, clip=true, totalheight=0.25\textheight, angle=90]{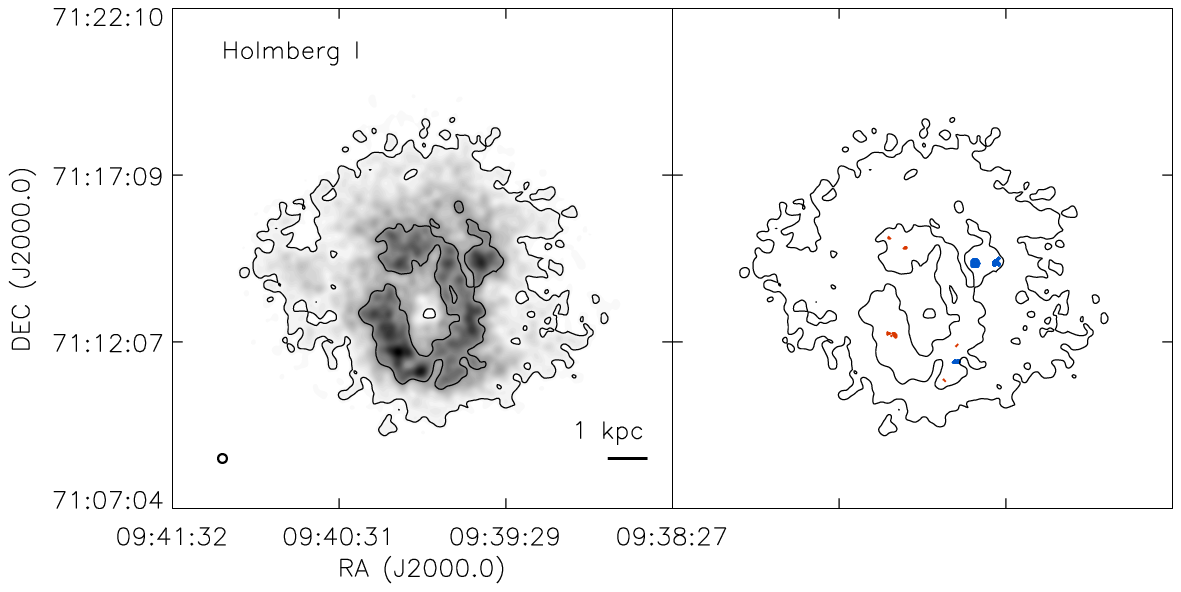} 
\includegraphics[trim=0cm 0cm 0cm 5.5cm, clip=true, totalheight=0.25\textheight, angle=90]{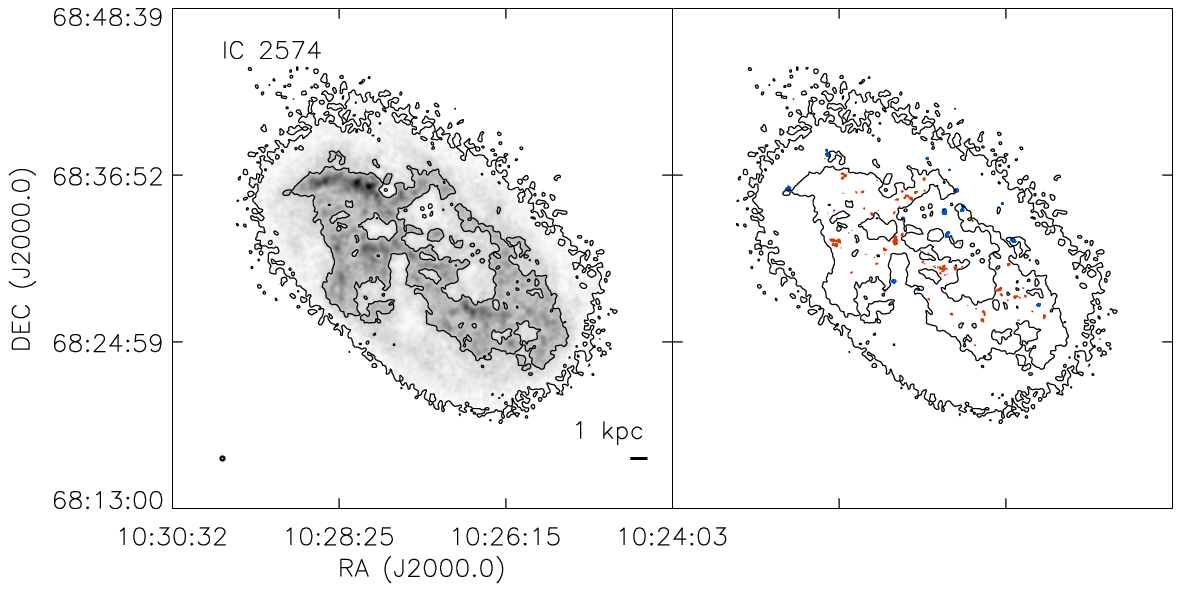}  
\includegraphics[trim=0cm 0cm 0cm 5.5cm, clip=true, totalheight=0.25\textheight, angle=90]{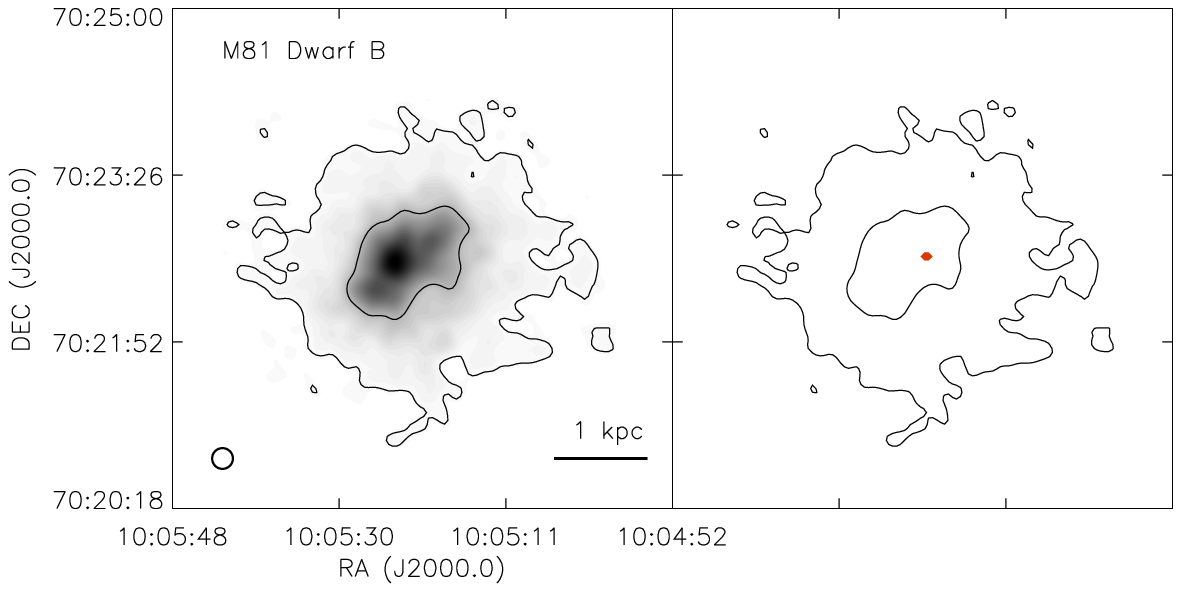} \\
\includegraphics[trim=0cm 0cm 0cm 5.5cm, clip=true, totalheight=0.25\textheight, angle=90]{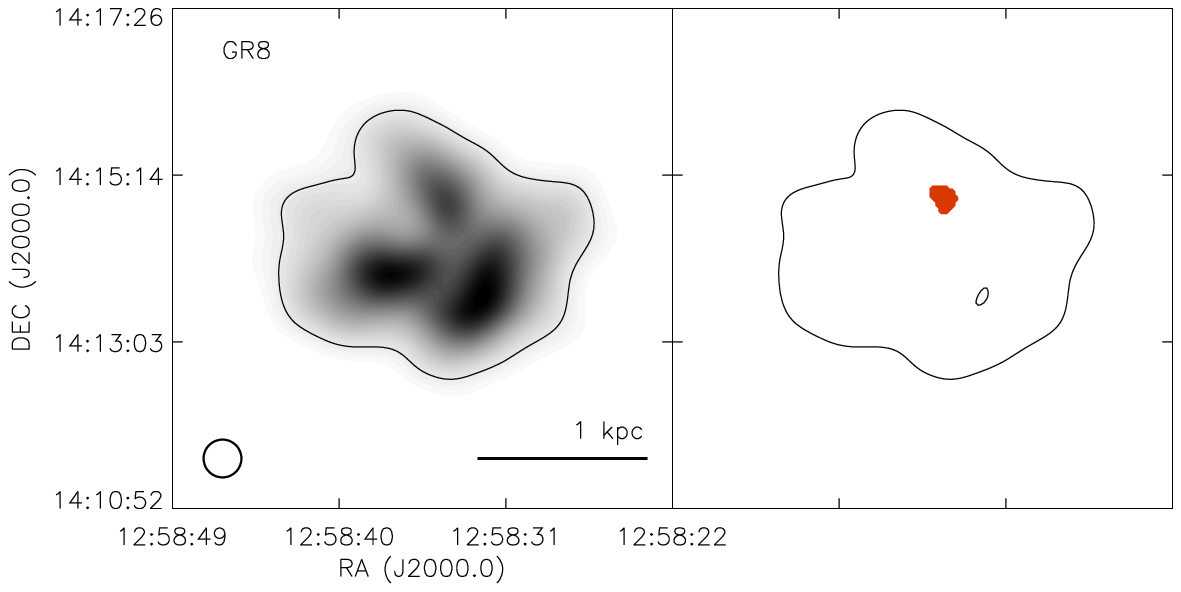} 
\includegraphics[trim=0cm 0cm 0cm 5.5cm, clip=true, totalheight=0.25\textheight, angle=90]{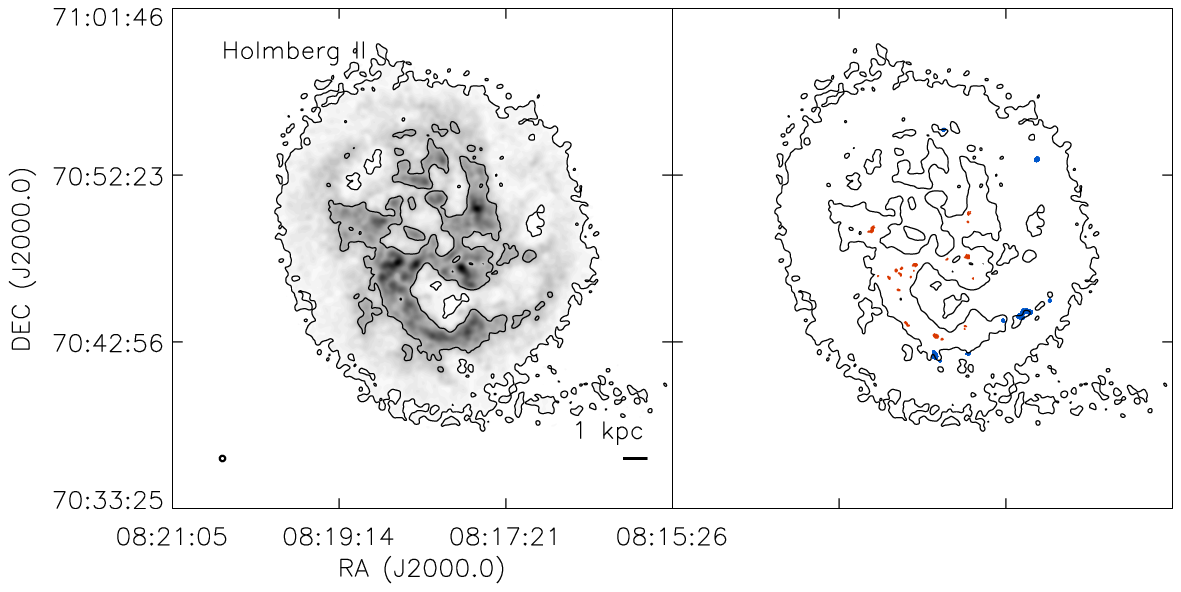}  
\includegraphics[trim=0cm 0cm 0cm 5.5cm, clip=true, totalheight=0.25\textheight, angle=90]{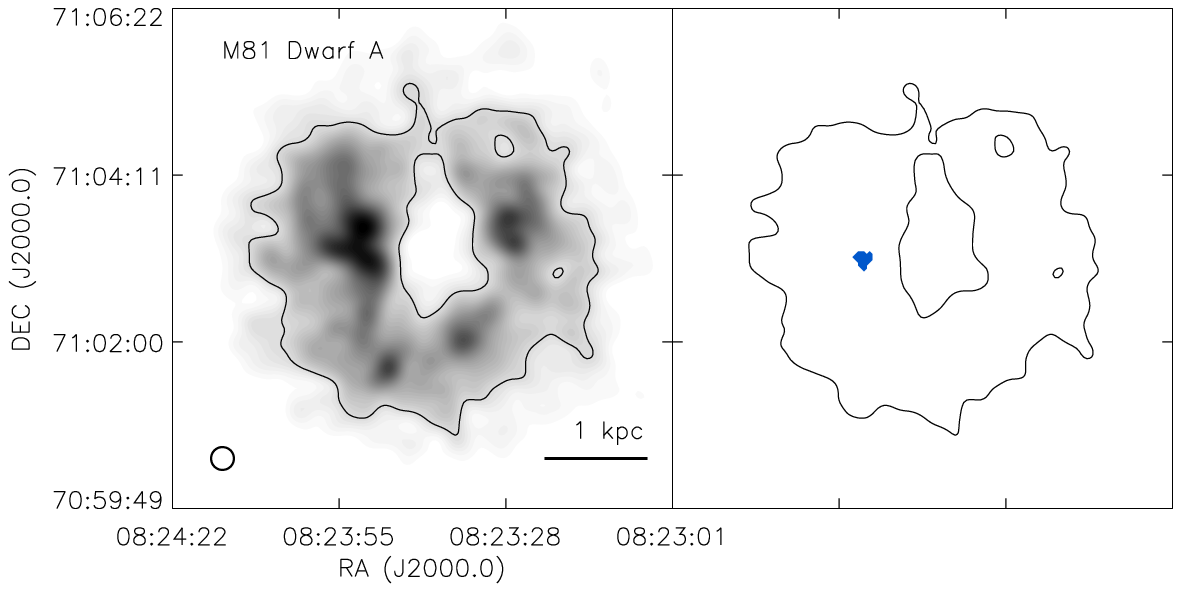}
\caption{Continued...}
\end{figure}

\clearpage

\begin{figure}
\figurenum{\ref{2plots}}
\includegraphics[trim=0cm 0cm 0cm 5.5cm, clip=true, totalheight=0.25\textheight, angle=90]{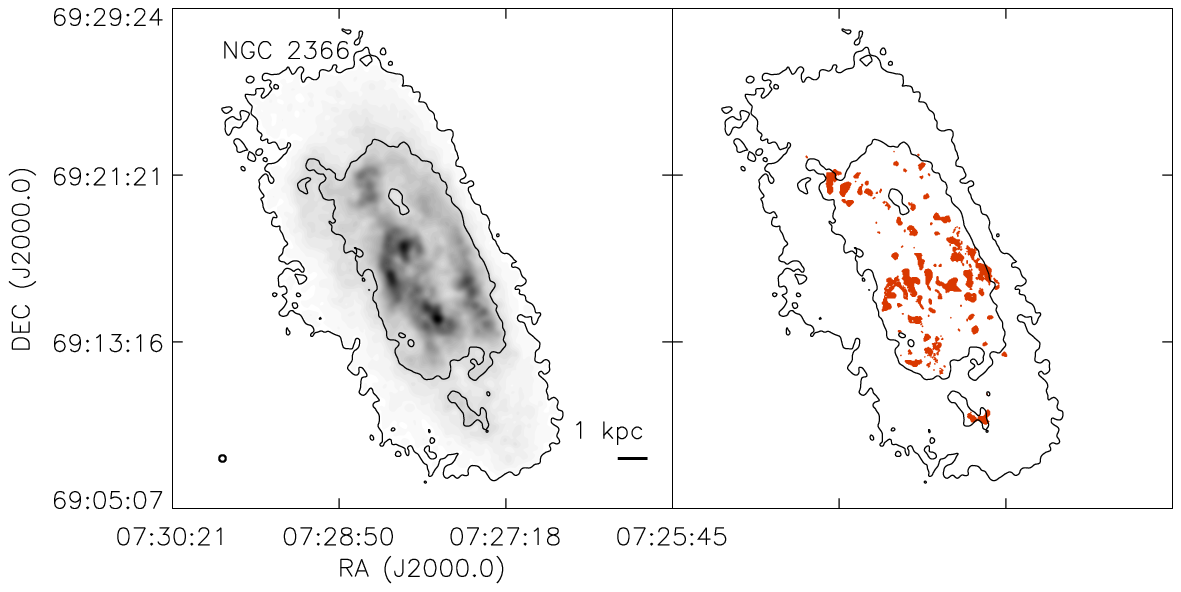} 
\includegraphics[trim=0cm 0cm 0cm 5.5cm, clip=true, totalheight=0.25\textheight, angle=90]{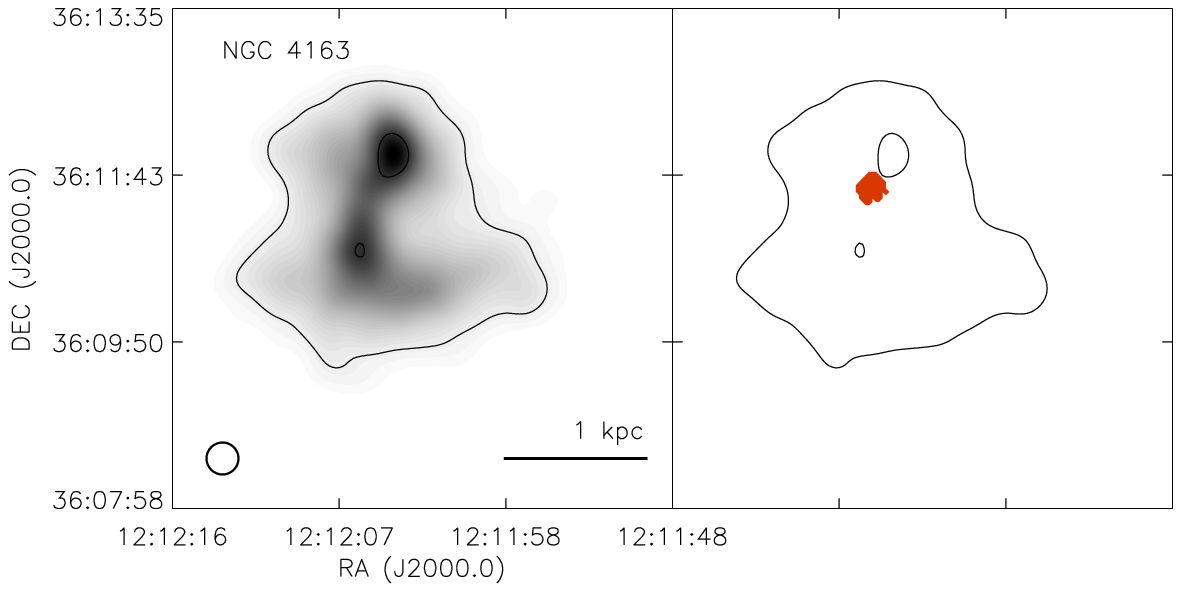}  
\includegraphics[trim=0cm 0cm 0cm 5.5cm, clip=true, totalheight=0.25\textheight, angle=90]{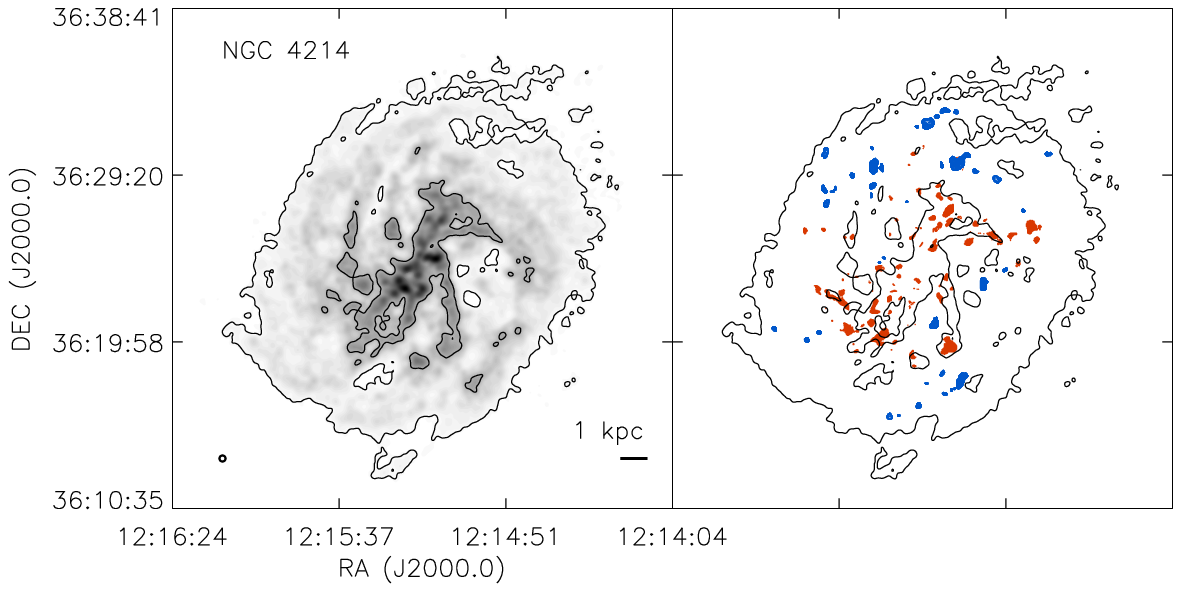} \\
\includegraphics[trim=0cm 0cm 0cm 5.5cm, clip=true, totalheight=0.25\textheight, angle=90]{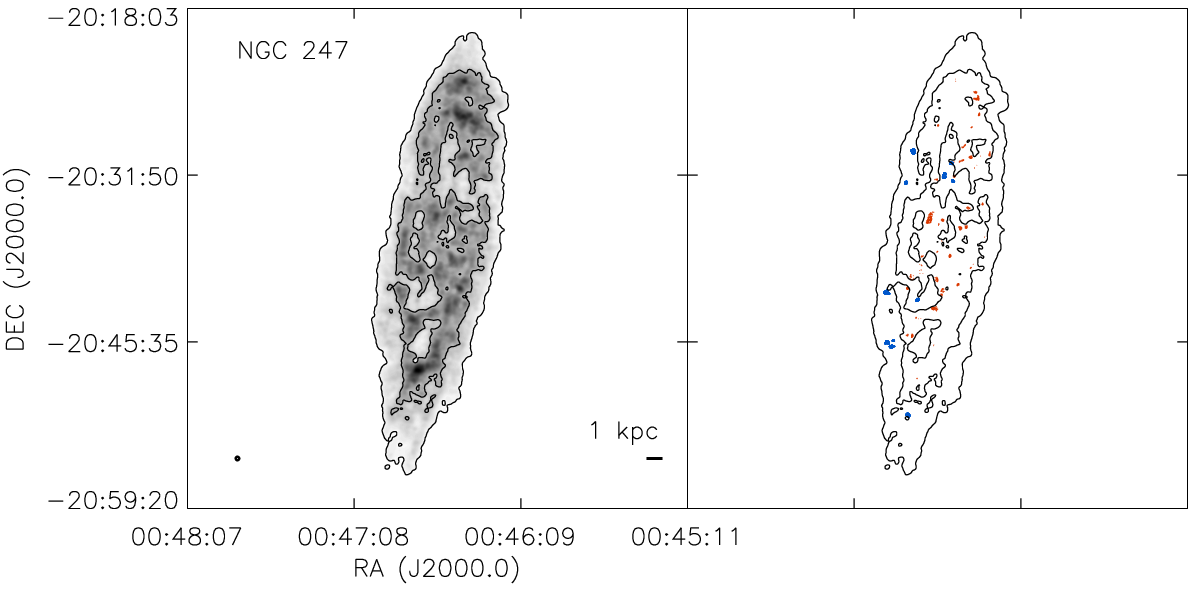} 
\includegraphics[trim=0cm 0cm 0cm 5.5cm, clip=true, totalheight=0.25\textheight, angle=90]{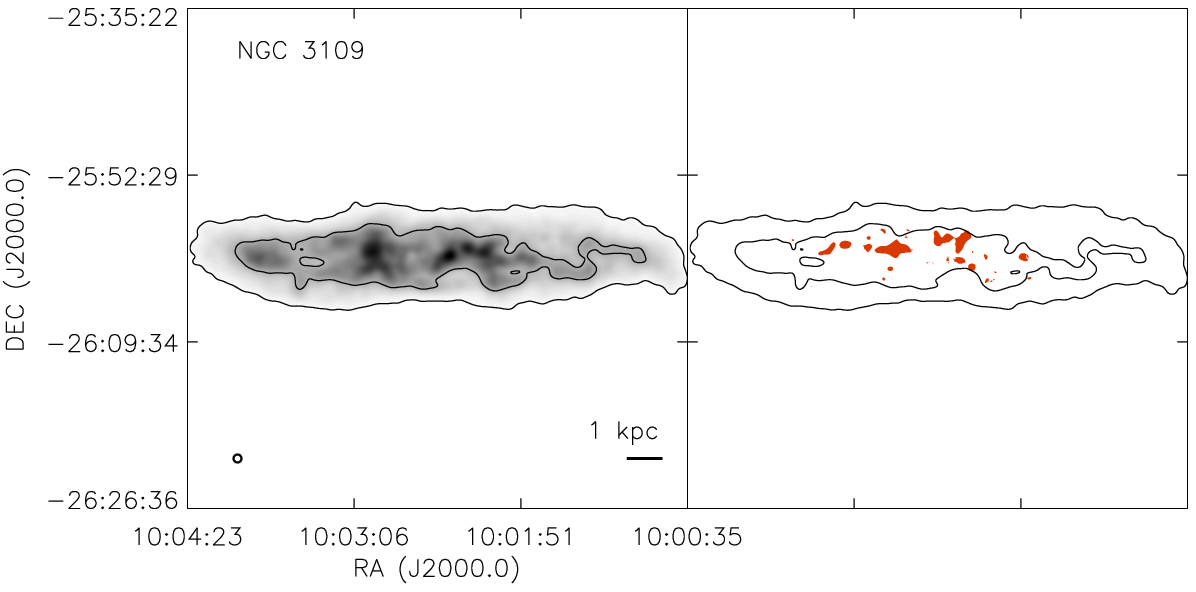}  
\includegraphics[trim=0cm 0cm 0cm 5.5cm, clip=true, totalheight=0.25\textheight, angle=90]{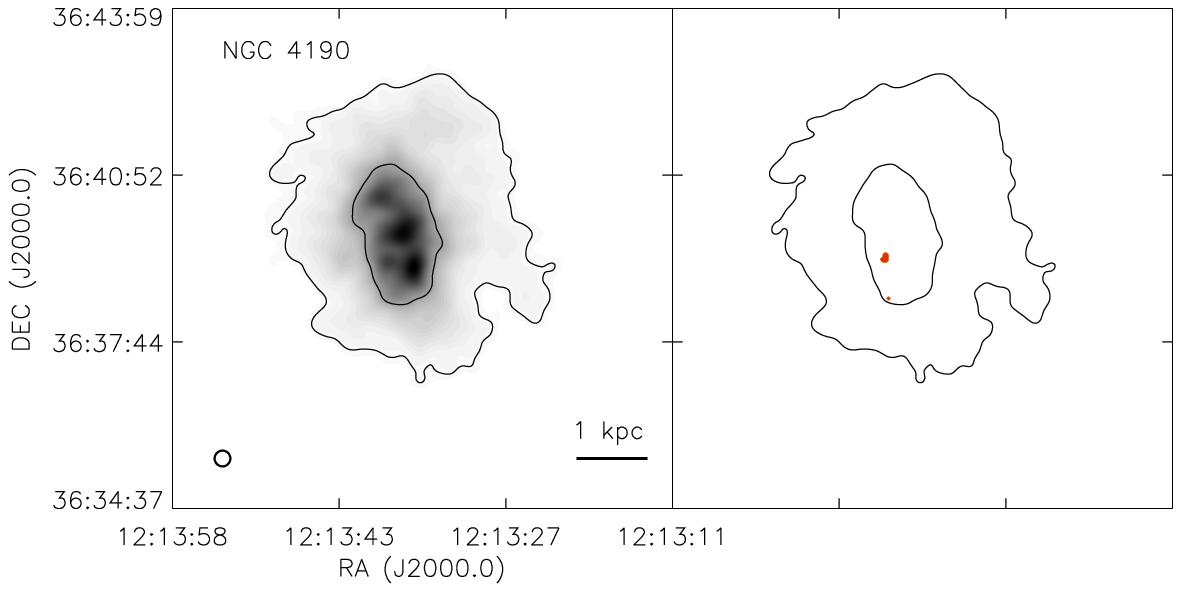}
\caption{Continued...}
\end{figure}

\clearpage

\begin{figure}
\figurenum{\ref{2plots}}
\includegraphics[trim=0cm 0cm 0cm 5.5cm, clip=true, totalheight=0.25\textheight, angle=90]{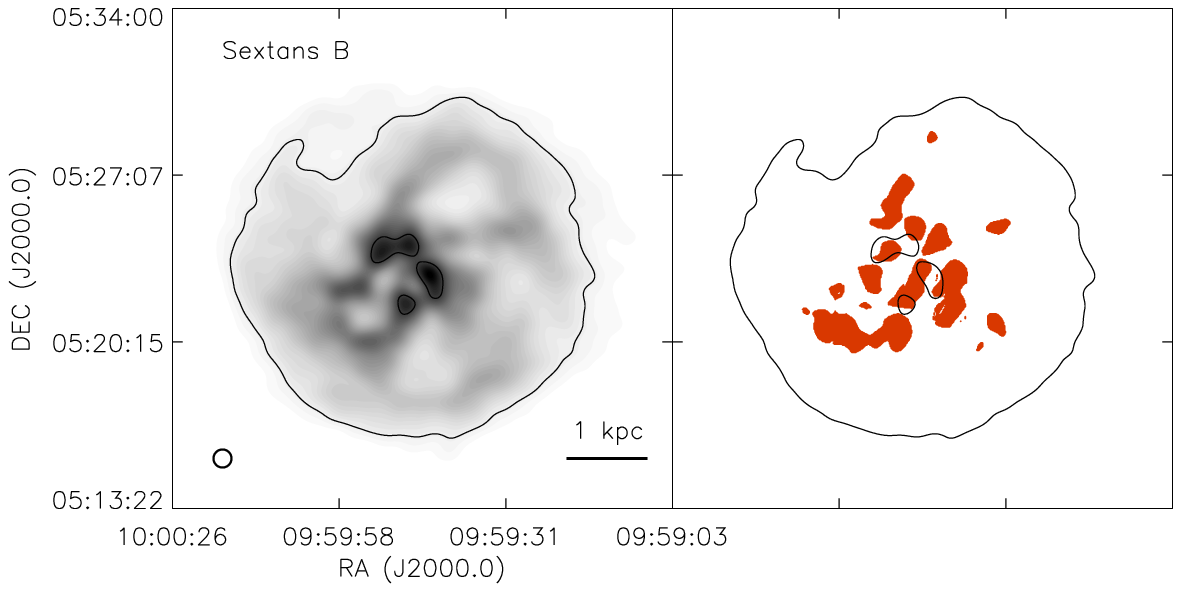}  
\includegraphics[trim=0cm 0cm 0cm 5.5cm, clip=true, totalheight=0.25\textheight, angle=90]{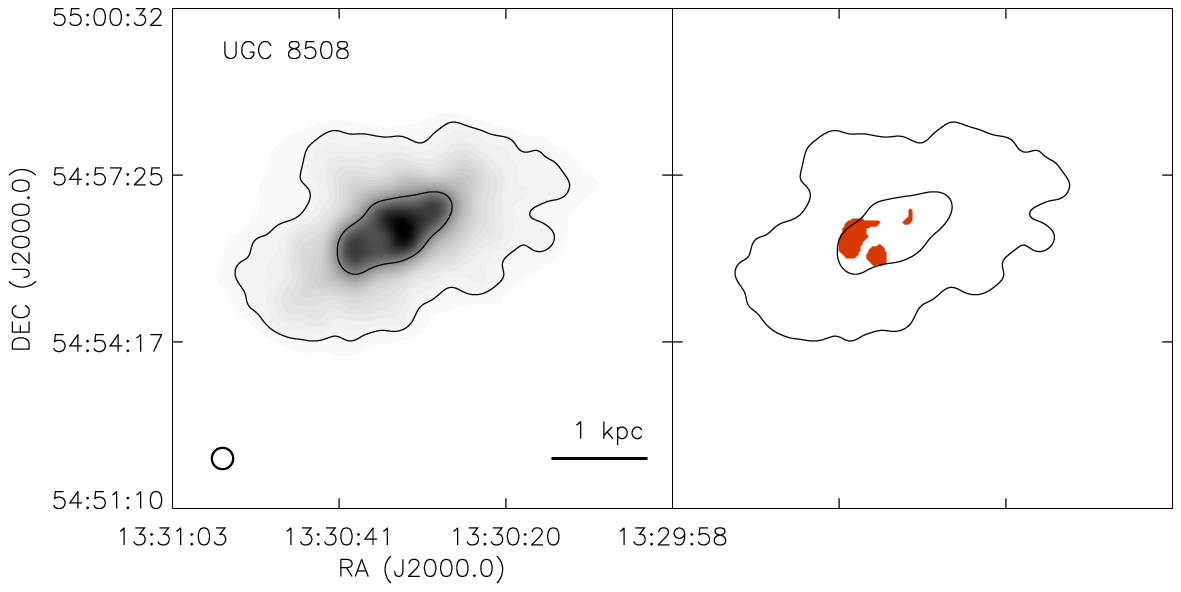} \\
\includegraphics[trim=0cm 0cm 0cm 5.5cm, clip=true, totalheight=0.25\textheight, angle=90]{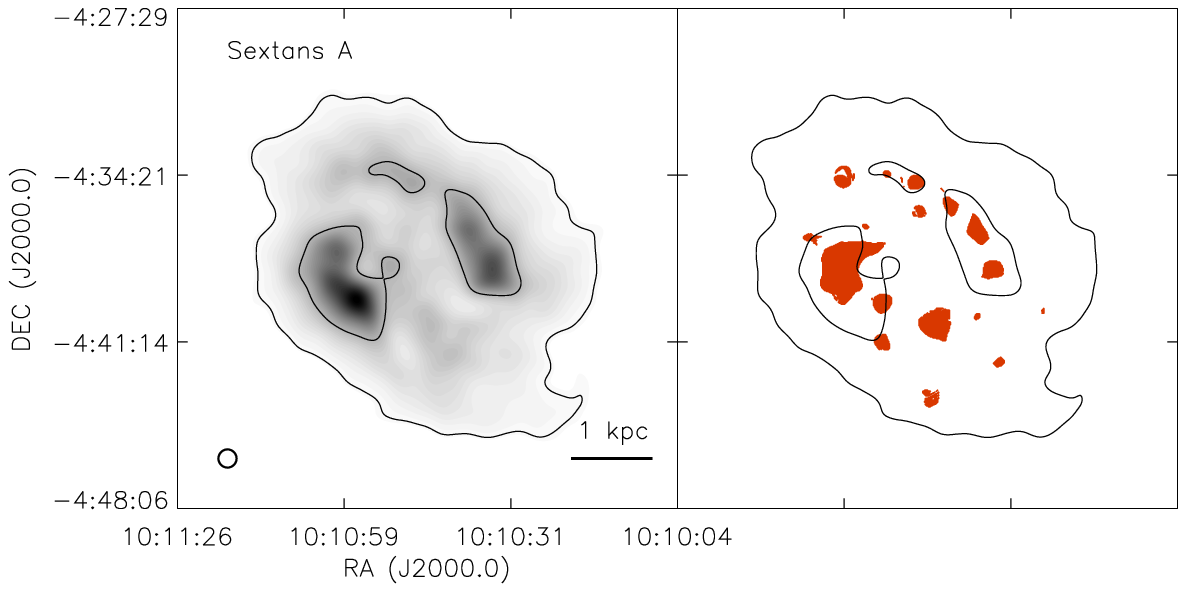} 
\includegraphics[trim=0cm 0cm 0cm 5.5cm, clip=true, totalheight=0.25\textheight, angle=90]{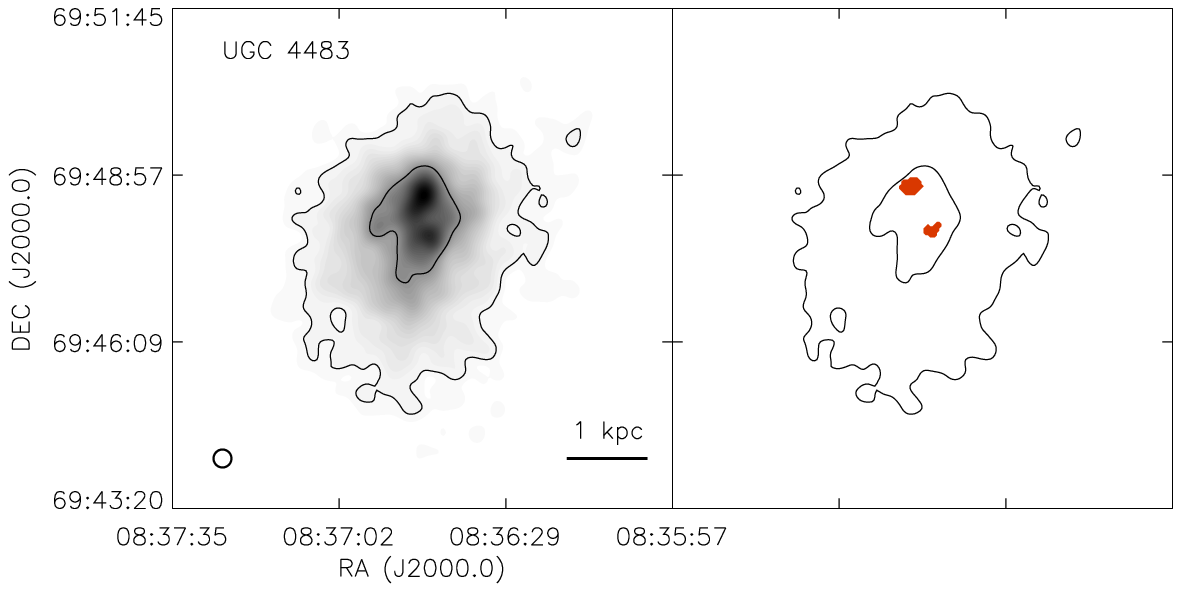}  
\includegraphics[trim=0cm 0cm 0cm 5.5cm, clip=true, totalheight=0.25\textheight, angle=90]{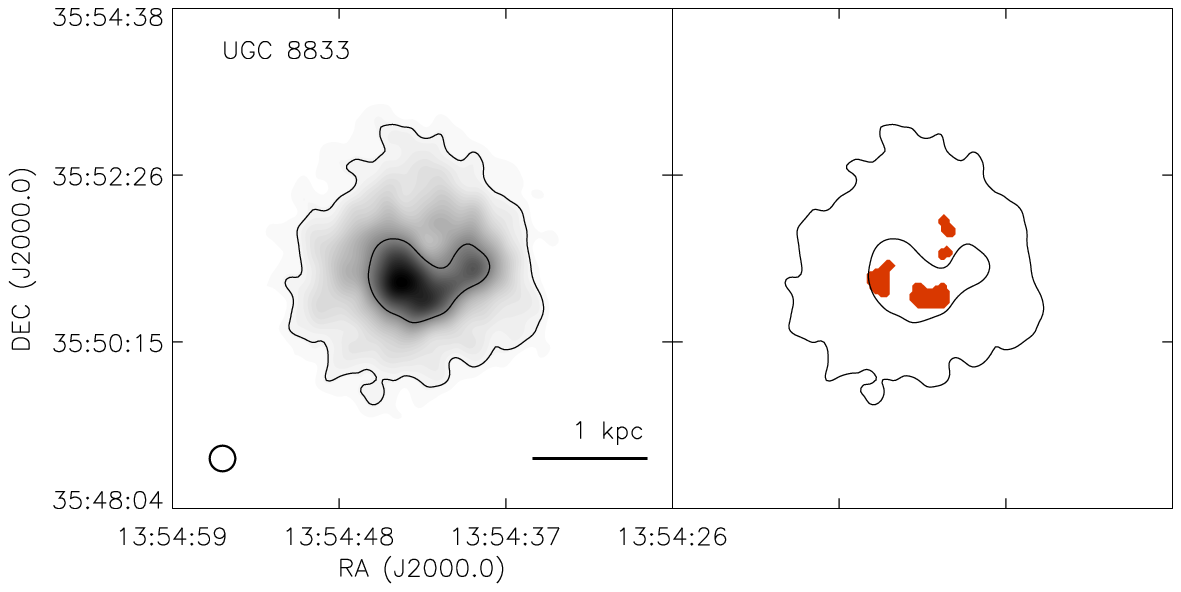}
\caption{Continued...}
\end{figure}

\clearpage

\clearpage

\clearpage

\clearpage

\clearpage

\clearpage

\clearpage

\clearpage

\clearpage

\end{document}